\documentclass{article}
\pdfoutput=1

\usepackage{arxiv}
\usepackage[utf8]{inputenc} 
\usepackage[T1]{fontenc}    
\usepackage{hyperref}       
\hypersetup{
colorlinks = true,
linkcolor = black,
anchorcolor = black,
citecolor = black,
filecolor = black,
urlcolor = blue
}
\usepackage{booktabs}       
\usepackage{amsfonts}       
\usepackage{nicefrac}       
\usepackage{microtype}      
\usepackage{amsmath,bm}
\usepackage{graphicx}
\usepackage[flushleft]{threeparttable}
\usepackage{multirow}
\usepackage{natbib}
\usepackage{pdflscape}
\usepackage{float}
\usepackage{placeins}
\usepackage[toc, page]{appendix}
\usepackage{lscape}
\usepackage{afterpage}
\usepackage{esvect}
\usepackage{amsmath}
\usepackage{ragged2e}
\justifying\let\raggedright\justifying
\usepackage{enumitem}
\usepackage{array}
\usepackage[table,xcdraw]{xcolor}

\newcommand{\figurehere}[1]{\begin{center}%
=========================\\%
Insert Figure #1 about here\\%
=========================\\%
\end{center}}
\newcommand{\tablehere}[1]{\begin{center}%
=========================\\%
Insert Table #1 about here\\%
=========================\\%
\end{center}}

\title{Two-step growth mixture model to examine heterogeneity in nonlinear trajectories}

\author{
    Jin Liu \thanks{CONTACT Jin Liu. Email: Veronica.Liu0206@gmail.com}\\
    Department of Biostatistics\\
    Virginia Commonwealth University \\
    \And
    Le Kang \\
    Department of Biostatistics\\
    Virginia Commonwealth University \\
    \And
    Roy T. Sabo \\
    Department of Biostatistics\\
    Virginia Commonwealth University \\
    \And
    Robert M. Kirkpatrick \\
    Department of Psychiatry \\
    Virginia Commonwealth University \\
    \And
    Robert A. Perera\\
    Department of Biostatistics\\
    Virginia Commonwealth University \\
}

\begin{document}
\maketitle

\begin{abstract}
Empirical researchers are usually interested in investigating the impacts that baseline covariates have when uncovering sample heterogeneity and separating samples into more homogeneous groups. However, a considerable number of studies in the structural equation modeling (SEM) framework usually start with vague hypotheses in terms of heterogeneity and possible causes. It suggests that (1) the determination and specification of a proper model with covariates is not straightforward, and (2) the exploration process may be computational intensive given that a model in the SEM framework is usually complicated and the pool of candidate covariates is usually huge in the psychological and educational domain where the SEM framework is widely employed. Following \citet{Bakk2017two}, this article presents a two-step growth mixture model (GMM) that examines the relationship between latent classes of nonlinear trajectories and baseline characteristics. Our simulation studies demonstrate that the proposed model is capable of clustering the nonlinear change patterns, and estimating the parameters of interest unbiasedly, precisely, as well as exhibiting appropriate confidence interval coverage. Considering the pool of candidate covariates is usually huge and highly correlated, this study also proposes implementing exploratory factor analysis (EFA) to reduce the dimension of covariate space. We illustrate how to use the hybrid method, the two-step GMM and EFA, to efficiently explore the heterogeneity of nonlinear trajectories of longitudinal mathematics achievement data.
\end{abstract}


\section{Introduction}\label{Intro}
\subsection{Motivating Example}
Earlier studies have examined the impacts of time-invariant covariates (TICs) have on nonlinear trajectories of mathematics achievement. For example, \citet{Liu2019BLSGM} associated nonlinear change patterns of mathematics IRT scaled scores to baseline covariates, including demographic information, socioeconomics factors, and school information. With the assumption that all covariates explain sample variability directly, this study showed that some baseline characteristics, such as sex, school type, family income, and parents' highest education can explain the heterogeneity in the nonlinear trajectories of mathematics scores. However, \citet{Kohli2015PLGC1} showed that latent classes of change patterns of mathematics achievement exist. Accordingly, these covariates may also inform latent class formation. In this study, we want to investigate the indirect impacts the covariates above and other baseline characteristics have on sample heterogeneity.

\subsection{Finite Mixture Model}
The finite mixture model (FMM) represents heterogeneity in a sample by allowing for a finite number of latent (unobserved) classes. The idea of mixture models is to put multiple probability distributions together in a sense using a linear combination. Although researchers may want to consider two different or multiple different families for the different kernels in some circumstances, the assumption that all latent classes' probability density functions follow normal distributions with class-specific parameters is common in application. 

This framework has gained considerable attention in the past twenty years among social and behavioral scientists due to its advantages over other clustering algorithms such as K-means for investigating sample heterogeneity. First, in the SEM framework, the FMM can incorporate any form of within-class models. For instance, \citet{Lubke2005FMM} specified factor mixture models, where the within-class model is a factor model to investigate heterogeneity in common factors, while \citet{Muthen1999GMM} defined growth mixture models (GMM), where the within-class model is a latent growth curve model to examine heterogeneity in trajectories. More importantly, the FMM is a model-based clustering method so that researchers can specify a model in this framework with domain knowledge: which parameters can be fixed to specific values, which need to be estimated, and which can be constrained to be equal (for example, invariance across classes). Additionally, the FMM is a probability-based clustering approach. Unlike other clustering methods, such as the K-means clustering algorithm, which aims to separate all observations into several clusters so that each entry belongs to one cluster without considering uncertainty, the FMM allows each element to belong to multiple classes at the same time with different posterior probabilities. 

This article focuses on the GMM with a nonlinear latent growth curve model as the within-class model. Specifically, trajectories in each class in the proposed GMM is a linear-linear piecewise model \citep{Harring2006nonlinear, Kohli2011PLGC, Kohli2013PLGC1, Kohli2013PLGC2, Sterba2014individually, Kohli2015PLGC1}), also referred to as a bilinear growth model \citep{Grimm2016growth, Liu2019knot, Liu2019BLSGM} with an unknown change-point (or knot). We decide to use the bilinear spline functional form as two existing studies, \citet{Kohli2015PLGC1} and \citet{Liu2019BLSGM}, have shown that a GMM with this functional form can capture the underlying change patterns of mathematics achievement and outperforms several parametric functions: linear, quadratic and Jenss-Bayley from the statistical perspective.

Similar to Liu et al. (2019), we propose the model in the framework of individual measurement occasions to account for possible heterogeneity in the measurement time in longitudinal studies \citep{Cook1983ITP, Mehta2000people, Finkel2003cognitive}. One possible solution to individual measurement occasions is to place the exact values of them to the matrix of factor loadings, which is termed the definition variable approach \citep{Mehta2000people, Mehta2005people}. Earlier studies have shown that the definition variable approach outperforms some approximate methods such as time-bins approach (where the assessment period is divided into several bins, and the factor loadings are set as those time-bins) in terms of bias, efficiency and Type I error rate \citep{Blozis2008coding, Coulombe2015ignoring}.

\subsection{Challenges of Finite Mixture Models Implementation}
A considerable number of studies in the SEM framework start from an exploratory stage where even empirical researchers only have vague assumptions about sample heterogeneity and its possible reasons. It suggests that we usually have two challenges when implementing a FMM, deciding the number of latent classes and selecting which covariates need to be included in the model. To investigate which criterion can be used to decide the number of latent classes, \citet{Nylund2007number} evaluated the performance of likelihood-based tests and the traditionally used information criteria, and showed that the bootstrap likelihood ratio test is a consistent indicator while the Bayesian information criterion (BIC) performs the best among all information criteria. Note that in practice, the BIC, which is calculated from the estimated likelihood directly, is usually more favorable due to its computational efficiency. 

It is also challenging to decide to include which covariates as predictors of class membership. Previous studies have shown that including subject-level predictors for latent classes can be realized by either one-step models \citep{Clogg1981one, Goodman1974one, Haberman1979one, Hagenaars1993one, Vermunt1997one, Bandeen1997one, Dayton1988one, Kamakura1994one, Yamaguchi2000one}, two-step models \citep{Bakk2017two} or three-step models \citep{Vermunt2010three, Bolck2004three, Asparouhov2014three}. The one-step model is suitable if a study is conducted in a confirmatory way or driven by answering a particular question, where specifying a proper mixture model for the covariates is usually a knowledge-driven process. On the contrary, the stepwise model is more suitable for an exploratory study in which empirical researchers usually have limited \textit{a priori} knowledge about possible class structure. For such studies, the current recommended approach is to investigate the number and nature of the latent classes without adding any covariates so that they do not inform class formation. In this study, we utilize the two-step model given that a considerable number of studies investigated by the SEM framework start from the exploratory stage and that \citet{Bakk2017two} has shown that the two-step procedure is consistently better than the three-step approach as it does not ignore the presence of uncertainty in the modal class assignments. Accordingly, by extending the method proposed in \citet{Bakk2017two} to the FMM with a bilinear spline growth curve as the within-class model, we first group nonlinear trajectories and estimate class-specific parameters with a pre-specified number of latent classes by fitting the measurement-model portion of the mixture model; we then investigate the associations between the latent classes and the individual-level covariates by fitting the measurement and structural model but fixing the measurement parameter estimates as their values from the first step. 

By utilizing the two-step model, we only need to refit the model in the second step rather than the whole model when adding or removing covariates, which saves the computational budget. However, the covariate space in the psychological and educational domains where the SEM framework is widely utilized is usually large, and there often exist subsets of highly correlated covariates. To address this issue, we propose to leverage a common multivariate data analysis approach in the SEM framework, exploratory factor analysis (EFA), to reduce the covariate space's dimension and address the issue of potential multicollinearity. Note that in this current study, it is not our aim to examine EFA comprehensively. We only want to demonstrate how to use the individual scores, for example, Thompson's scores \citep{Thomson1939score} or Bartlett's weighted least-squares scores \citep{Bartlett1937score}, based on the output of EFA, with a basic understanding of its algorithm.

EFA is a useful multivariate data analysis approach to explain the variance-covariance matrix of the dataset by replacing a large set of manifest variables with a smaller set of latent variables. In this approach, manifested variables are assumed to be caused by latent variables. When implementing EFA, we impose no constraints on the relationships between manifested and latent variables. With an assumption that all manifested variables are related to all latent variables, this approach aims to determine the appropriate number of factors and factor loadings (i.e., correlations between observed variables and unobserved variables). Next, we calculate a score for each factor of each individual based on the factor loadings and standardized covariate values. We then view these individual-level scores instead of the covariates as baseline characteristics in the second step.

The proposed hybrid mothed aims to provide an analytical framework for examining heterogeneity in an exploratory study. We extend the two-step method proposed by \citet{Bakk2017two} to investigate the heterogeneity in nonlinear trajectories in the framework of individually varying time points (ITPs). Specifically, we consider the bilinear spline growth curve with an unknown knot, which can capture the shape of trajectories of our motivating data shown in earlier studies \citep{Kohli2015PLGC1, Liu2019BLSGM}, as the functional form of the underlying change patterns. We specify the model with truly individually measurement occasions, which are ubiquity in longitudinal studies, to avoid unnecessary inadmissible estimation. Additionally, we propose to use EFA to reduce the dimension of the covariate space. 

The remainder of this article is organized as follows. We describe the model specification and model estimation of the two-step growth mixture model in the framework of ITPs in the method section. In the subsequent section, we describe the design of the Monte Carlo simulation for model evaluation. We evaluate the model performance through the performance measures, which include the relative bias, the empirical standard error (SE), the relative root-mean-squared-error (RMSE), and the empirical coverage for a nominal $95\%$ confidence interval of each parameter of interest, as well as the accuracy. We then introduce the dataset of repeated mathematics achievement scores (mathematics IRT scaled score) from the Early Childhood Longitudinal Study, Kindergarten Class of 2010-11 (ECLS-K: 2011), and demonstrate the hybrid method's implementation in the application section. Finally, discussions are framed concerning the model's limitations as well as future directions.

\section{Method}\label{Method}
\subsection{Model Specification}\label{M:Specify}
In this section, we specify the GMM with a linear-linear piecewise growth curve with an unknown knot as a within-class model. \citet{Harring2006nonlinear} pointed out there are five parameters in the bilinear spline functional form: an intercept and slope of each linear piece and a change-point, yet the degree-of-freedom of the bilinear spline is four since two linear pieces join at the knot. In this study, we view the initial status, two slopes and the knot as the four parameters. We construct the model with an assumption that the class-specific knot is the same across all individuals in a latent class though \citet{Preacher2015repara, Liu2019BLSGM} have shown that the knot can be an additional growth factor by relaxing the assumption. Suppose the pre-specified number of latent classes is $K$, for $i=1$ to $n$ individuals and $k=1$ to $K$ latent classes, we express the model as

\begin{align}
&p(\boldsymbol{y}_{i}|z_{i}=k,\boldsymbol{x}_{i})=\sum_{k=1}^{K}\pi(z_{i}=k|\boldsymbol{x}_{i})\times p(\boldsymbol{y}_{i}|z_{i}=k),\label{eq:GMM}\\
&\pi(z_{i}=k|\boldsymbol{x}_{i})=\begin{cases}
\frac{1}{1+\sum_{k=2}^{K}\exp(\beta_{0}^{(k)}+\boldsymbol{\beta}^{(k)T}\boldsymbol{x}_{i})}& \text{Reference Group ($k=1$)}\\
\frac{\exp(\beta_{0}^{(k)}+\boldsymbol{\beta}^{(k)T}\boldsymbol{x}_{i})} {1+\sum_{k=2}^{K}\exp(\beta_{0}^{(k)}+\boldsymbol{\beta}^{(k)T}\boldsymbol{x}_{i})} & \text{Other Groups ($k=2,\dots, K$)}
\end{cases},\label{eq:step2}\\
&\boldsymbol{y}_{i}|(z_{i}=k)=\boldsymbol{\Lambda}_{i}\boldsymbol{\eta}_{i}+\boldsymbol{\epsilon}_{i},\label{eq:step1_1}\\
&\boldsymbol{\eta}_{i}|(z_{i}=k)=\boldsymbol{\mu_{\eta}}^{(k)}+\boldsymbol{\zeta}_{i}.\label{eq:step1_2}
\end{align}
Equation (\ref{eq:GMM}) defines a FMM that combines mixing proportions, $\pi(z_{i}=k|\boldsymbol{x}_{i})$, and within-class models, $p(\boldsymbol{y}_{i}|z_{i}=k)$, where $\boldsymbol{x}_{i}$, $\boldsymbol{y}_{i}$ and $z_{i}$ are the covariates, $J\times1$ vector of repeated outcome (in which $J$ in the number of measurements) and membership of the $i^{th}$ individual, respectively. For Equation (\ref{eq:GMM}), we have two constratints: $0\le \pi(z_{i}=k|\boldsymbol{x}_{i})\le 1$ and $\sum_{k=1}^{K}\pi(z_{i}=k|\boldsymbol{x}_{i})=1$. Equation (\ref{eq:step2}) defines mixing components as logistic functions of covariates $\boldsymbol{x}_{i}$, where $\beta_{0}^{(k)}$ and $\boldsymbol{\beta}^{(k)}$ are logistic coefficients. These functions decide the membership for the $i^{th}$ individual, depending on the values of the covariates $\boldsymbol{x}_{i}$. 

Equations (\ref{eq:step1_1}) and (\ref{eq:step1_2}) together define a within-class model. Similar to all factor models, Equation (\ref{eq:step1_1}) expresses the outcome $\boldsymbol{y}_{i}$ as a linear combination of growth factors. When the underlying functional form is bilinear spline growth curve with an unknown knot without considering variability, $\boldsymbol{\eta}_{i}$ is a $3\times1$ vector of growth factors ($\boldsymbol{\eta}_{i}=\eta_{0i},\eta_{1i},\eta_{2i}$, for an initial status and a slope of each stage of the $i^{th}$ individual). Accordingly, $\boldsymbol{\Lambda}_{i}$, which is a function of the class-specific knot $\gamma^{(k)}$, is a $J\times3$ matrix of factor loadings. The pre- and post-knot $\boldsymbol{y}_{i}$ can be expressed as 
\begin{equation}\nonumber
y_{ij}=\begin{cases}
\eta_{0i}+\eta_{1i}t_{ij}+\epsilon_{ij} & t_{ij}\le\gamma^{(k)}\\
\eta_{0i}+\eta_{1i}\gamma^{(k)}+\eta_{2i}(t_{ij}-\gamma^{(k)})+\epsilon_{ij} & t_{ij}>\gamma^{(k)}\\
\end{cases}, 
\end{equation}
where $y_{ij}$ and $t_{ij}$ are the measurement and measurement occasion of the $i^{th}$ individual at time $j$. Additionally, $\boldsymbol{\epsilon}_{i}$ is a $J\times 1$ vector of residuals of the $i^{th}$ individual. Equation (\ref{eq:step1_2}) further expresses the growth factors as deviations from their class-specific means. In the equation, $\boldsymbol{\mu_{\eta}}^{(k)}$ is a $3\times 1$ vector of class-specific growth factor means and $\boldsymbol{\zeta}_{i}$ is a $3\times 1$ vector of residual deviations from the mean vector of the $i^{th}$ individual.

To unify pre- and post-knot expressions, we need to reparameterize growth factors. Earlier studies (for example, \citet{Harring2006nonlinear, Grimm2016growth, Liu2019BLSGM}) presented multiple ways to realize this aim. Note that no matter which approach we follow to reparameterize growth factors, the reparameterized coefficients are not directly relevant to the underlying change patterns and then need to be transformed back to be interpretable. In this article, we follow the reparameterized method in \citet{Grimm2016growth} and define the class-specific reparameterized growth factors as the measurement at the knot, the mean of two slopes and the half difference of two slopes. Note that the expressions of the repeated outcome $\boldsymbol{y}_{i}$ using the growth factors in the original and reparameterized frames are equivalent. We also extend the (inverse-)transformation functions and matrices for the reduced model in \citet{Liu2019BLSGM}, with which we can obtain the original parameters efficiently for interpretation purposes. Detailed class-specific reparameterizing process and the class-specific (inverse-) transformation are provided in Appendix \ref{Supp:1A} and Appendix \ref{Supp:1B}, respectively. 

\subsection{Model Estimation}
To simplify the model, we assume that class-specific growth factors follow a multivariate Gaussian distribution, that is, the vector of deviations $\boldsymbol{\zeta}_{i}|k\sim \text{MVN}(\boldsymbol{0}, \boldsymbol{\Psi_{\eta}}^{(k)})$. Note that $\boldsymbol{\Psi_{\eta}}^{(k)}$ is a $3\times 3$ variance-covariance matrix of class-specific growth factors. We also assume that individual residuals follow identical and independent normal distributions over time in each latent class, that is, $\boldsymbol{\epsilon}_{i}|k\sim N(\boldsymbol{0}, \theta_{\epsilon}^{(k)}\boldsymbol{I})$, where $\boldsymbol{I}$ is a $J\times J$ identity matrix. Accordingly, for the $i^{th}$ individual in the $k^{th}$ unobserved group, the within-class model implied mean vector ($\boldsymbol{\mu}_{i}^{(k)}$) and variance-covariance matrix of repeated measurements ($\boldsymbol{\Sigma}_{i}^{(k)}$) are
\begin{align}
&\boldsymbol{\mu}_{i}^{(k)}=\boldsymbol{\Lambda}_{i}\boldsymbol{\mu_{\eta}}^{(k)},\label{eq:parameter1.1.1}\\
&\boldsymbol{\Sigma}_{i}^{(k)}=\boldsymbol{\Lambda}_{i}\boldsymbol{\Psi_{\eta}}^{(k)}\boldsymbol{\Lambda}_{i}^{T}+\theta_{\epsilon}^{(k)}\boldsymbol{I}.\label{eq:parameter1.1.2}
\end{align} 

\subsubsection*{Step 1}
In the first step, we estimate the class-specific parameters and mixing proportions for the model specified in Equations (\ref{eq:GMM}), (\ref{eq:step2}), (\ref{eq:step1_1}) and (\ref{eq:step1_2}) without considering the impact of covariates $\boldsymbol{x}_{i}$ have on the class formation. The parameters need to be estimated in this step include
\begin{equation}\nonumber
\boldsymbol{\Theta}_{s1}=\{\mu_{\eta_{0}}^{(k)}, \mu_{\eta_{1}}^{(k)}, \mu_{\eta_{2}}^{(k)}, \gamma^{(k)}, \psi_{00}^{(k)}, \psi_{01}^{(k)}, \psi_{02}^{(k)}, \psi_{11}^{(k)}, \psi_{12}^{(k)}, \psi_{22}^{(k)}, \theta_{\epsilon}^{(k)}, \pi^{(2)}, \cdots, \pi^{(K)}\}.
\end{equation}
We employ full information maximum likelihood (FIML) technique, which accounts for the potential heterogeneity of individual contributions to the likelihood, to estimate $\boldsymbol{\Theta}_{s1}$. The log-likelihood function of the model specified in Equations (\ref{eq:GMM}), (\ref{eq:step2}), (\ref{eq:step1_1}) and (\ref{eq:step1_2}) without the effect of $\boldsymbol{x}_{i}$ is
\begin{equation}\label{eq:lik}
\begin{aligned}
\ell\ell(\boldsymbol{\Theta}_{\text{s1}})&=\sum_{i=1}^{n}\log\bigg(\sum_{k=1}^{K}\pi(z_{i}=k)p(\boldsymbol{y}_{i}|z_{i}=k)\bigg)\\
&=\sum_{i=1}^{n}\log\bigg(\sum_{k=1}^{K}\pi(z_{i}=k)p(\boldsymbol{y}_{i}|\boldsymbol{\mu}_{i}^{(k)},\boldsymbol{\Sigma}_{i}^{(k)})\bigg).
\end{aligned}
\end{equation}

\subsubsection*{Step 2}
In the second step, we examine the associations between the latent classes and the individual-level covariates by fixing the class-specific parameters' estimates as their values from the first step, that is, the parameters need to be estimated in this step are those logistic coefficients, $\boldsymbol{\Theta}_{\text{s2}}=\{\beta_{0}^{(k)}, \boldsymbol{\beta}^{T(k)}\boldsymbol{}\}\ (k=2,\dots,K$), in Equation (\ref{eq:step2}). The log-likelihood function in Equation (\ref{eq:lik}) is also need to be modified as
\begin{equation}
\begin{aligned}
\ell\ell(\boldsymbol{\Theta}_{\text{s2}})&=\sum_{i=1}^{n}\log\bigg(\sum_{k=1}^{K}\pi(z_{i}=k| \boldsymbol{x}_{i})p(\boldsymbol{y}_{i}|z_{i}=k)\bigg)\\
&=\sum_{i=1}^{n}\log\bigg(\sum_{k=1}^{K}\pi(z_{i}=k| \boldsymbol{x}_{i})p(\boldsymbol{y}_{i}|\boldsymbol{\hat{\mu}}_{i}^{(k)},\boldsymbol{\hat{\Sigma}}_{i}^{(k)})\bigg).
\end{aligned}
\end{equation}

We construct the proposed two-step GMM using the R package \textit{OpenMx} with the optimizer CSOLNP \citep{Pritikin2015OpenMx, OpenMx2016package, User2018OpenMx, Hunter2018OpenMx}, with which we can fit the proposed GMM and implement the class-specific inverse-transformation matrices to obtain coefficients that are directly related to underlying change patterns as shown in Appendix \ref{Supp:1B}. In the online appendix (\url{https://github.com/Veronica0206/Dissertation_projects}), we provide the \textit{OpenMx} code for the proposed model as well as a demonstration. For the researchers interested in using \textit{Mplus}, we also provide \textit{Mplus} 8 code for the model in the online appendix. 

\section{Model Evaluation}
We evaluate the proposed model using a Monte Carlo simulation study with two goals. The first goal is to evaluate the model performance by examining the relative bias, empirical SE, relative RMSE and empirical coverage for a nominal 95\% confidence interval (CI) of each parameter. Table \ref{tbl:metric} lists the definitions and estimates of these performance metrics. The second goal is to evaluate the accuracy, which is defined as the fraction of all correctly classified instances \citep{Bishop2006pattern} as we have true labels in simulation studies. To obtain the accuracy, we need to calculate the posterior probabilities for each individual belonging to the $k^{th}$ unobserved group based on the estimates obtained from the first step by Bayes' theorem
\begin{equation}\nonumber
p(z_{i}=k|\boldsymbol{y}_{i})=\frac{\pi(z_{i}=k)p(\boldsymbol{y}_{i}|z_{i}=k)}{\sum_{k=1}^{K}\pi(z_{i}=k)p(\boldsymbol{y}_{i}|z_{i}=k)}
\end{equation}
and assign each individual to the latent class with the highest posterior probability to which that observation most likely belongs. If multiple posterior probabilities equal to the maximum value, we break the tie among competing components randomly \citep{McLachlan2000FMM}.

\tablehere{1}

We decided the number of repetitions $S=1,000$ by an empirical approach proposed by \citet{Morris2019simulation} in the simulation design. The (relative) bias is the most important performance metric in our simulation, so we want to keep its Monte Carlo standard error\footnote{$\text{Monte Carlo SE(Bias)}=\sqrt{Var(\hat{\theta})/S}$ \citep{Morris2019simulation}.} less than $0.005$. We ran a pilot simulation study and noted that standard errors of all parameters except the intercept variances were less than $0.15$, so we needed at least $900$ replications to make sure the Monte Carlo standard error of bias is as low as we expected. We then decided to proceed with $S=1,000$ to be more conservative. 

\subsection{Design of Simulation Study}
The simulation study has two parts. As mentioned earlier, we propose the two-step model with a bilinear spline growth curve with an unknown knot as the within-class model assuming that the change-point is roughly similar for all individuals in each latent class as the knot variance is not the primary interest of this study. In the first part, we restricted the knot to be identical for all trajectories in a latent class to evaluate the model performance when being specified correctly. We are also interested in examining how the proposed model works when relaxing the restriction. Accordingly, in the second part, by allowing for the individual difference in the knot, we investigated the robustness of the proposed model by assessing the model performance in the presence of knots with the standard deviation set as $0.3$. 

We list all conditions of simulation studies for Part 1 and Part 2 in Table \ref{tbl:simu}. All conditions except the knot variance for both parts were set to be the same. For both parts, we fixed the conditions that are not of the primary interests of the current study. For example, we considered ten scaled and equally spaced waves since \citet{Liu2019BLSGM} has shown that the bilinear growth model had decent performance concerning the performance measures when being applied to a longitudinal data set with ten repeated measures and fewer number of repeated measures did not affect model performance meaningfully. Similar to \citet{Liu2019BLSGM}, we allowed the time-window of individual measurement occasions ranging from $-0.25$ and $+0.25$, which was viewed as a `medium' deviation, as an existing simulation study \citep{Coulombe2015ignoring}, around each wave. We also fixed the variance-covariance matrix of the class-specific growth factors that usually change with the time scale and the measurement scale in practice; accordingly, we simply kept the index of dispersion ($\sigma^{2}/\mu$) of each growth factor at a tenth scale, guided by \citet{Bauer2003GMM, Kohli2011PLGC, Kohli2015PLGC1}. Further, the growth factors were set to be positively correlated to a moderate degree ($\rho=0.3$). 

\tablehere{2}

For both parts, the primary aim was to investigate how the separation between latent classes, the unbalanced class mixing proportion, and the trajectory shape affected the model performance. Utilizing a model-based clustering algorithm, we are usually interested in examining how well the model can detect heterogeneity in samples and estimate parameters of interest in each latent class. Intuitively, the model should perform better under those conditions with larger separation between latent classes. We wanted to test this hypothesis. In the simulation design, we had two metrics to gauge the separation between clusters: the difference between the knot locations and the Mahalanobis distance (MD) of the three growth factors of latent classes. We set $1$, $1.5$ and $2$ as a small, medium, and large difference between the knot locations. We chose $1$ as the level of small difference to follow the rationale in \citet{Kohli2015PLGC1} and considered the other two levels to investigate whether the more widely spaced knots improve the model performance. We considered two levels of MD, $0.86$ (i.e., small distance) and $1.72$ (i.e., large distance), for class separation. Note that both the small and large distance in the current simulation design was smaller than the corresponding level in \citet{Kohli2015PLGC1} because we want to examine the proposed model under more challenging conditions in terms of cluster separation. 

We chose two levels of mixing proportion, $1:1$ and $1:2$, for the conditions with two latent classes and three levels of mixing proportion, $1:1:1$, $1:1:2$ and $1:2:2$, for the scenarios with three clusters. We selected these levels because we wanted to evaluate how the challenging conditions (i.e., the unbalanced allocation) affect performance measures and the accuracy. We also examined several common change patterns shown in Table \ref{tbl:simu} (Scenario 1, 2 and 3). Under each scenario, we changed the knot locations and one growth factor but fixed the other two growth factors to satisfy the specified MD. We considered $\theta=1$ or $\theta=2$ as two levels of homogeneous residual variances across latent classes to see the effect of the measurement precision, and we considered two levels of sample size. 

\subsection{Label Switching}
All mixture models suffer from the label switching issue, that is, inconsistent assignments of membership for multiple replications in simulation studies. The label switching does not hurt the model estimation in the frequentist framework since the likelihood is invariant to permutation of cluster labels; however, the estimates from the first latent class may be mislabeled as such from other latent classes (Class 2 or Class 3 in our case) \citep{Tueller2011labels}. In this study, we utilized the column maxima switched label detection algorithm developed by \citet{Tueller2011labels} to check whether the labels were switched; and if it happens, the final estimates were relabeled in the correct order before model evaluation.

\subsection{Data Generation and Simulation Step}\label{E:step}
For each condition listed in Table \ref{tbl:simu}, we used two-step data generation to obtain a component label $z_{i}$ for each individual and then generated data for each component. The general steps of the simulation for the proposed two-step model in the framework of individual measurement occasions were carried out as follows:
\begin{enumerate}
\item Created component label $z_{i}$ for the $i^{th}$ individual:
\begin{enumerate}
\item Generated data matrix of exogenous variables,
\item Calculated the probability vector for each entry with a set of specified regression coefficients using a multinomial logit link and assigned a component label $z_{i}$ to each observation,
\end{enumerate} 
\item Generated data for growth factors and a knot of each latent class using the R package \textit{MASS} \citep{Venables2002Statistics},
\item Generated the time structure with $J$ scaled and equally-spaced waves $t_{j}$ and obtained individual measurement occasions: $t_{ij}\sim U(t_{j}-\Delta, t_{j}+\Delta)$ by allowing disturbances around each wave,
\item Calculated factor loadings, which are functions of ITPs and the knot, for each individual,
\item Calculated values of the repeated measurements based on the class-specific growth factors, corresponding factor loadings, and residual variances,
\item Applied the proposed model to the generated data set, estimated the parameters, and constructed corresponding $95\%$ Wald CIs, as well as calculated posterior probabilities that each individual belongs to each of the multiple latent classes, followed by accuracy,
\item Repeated the above steps until after obtaining $1,000$ convergent solutions to calculate the mean accuracy, performed the column maxima switched label detection algorithm, relabeled (if labels had been switched), and calculated the relative bias, empirical SE, relative RMSE and coverage probability of each parameter under investigation.
\end{enumerate}

\section{Result}
\subsection{Model Convergence}
In this section, we first examined the convergence\footnote{In our project, convergence is defined as to reach \textit{OpenMx} status code $0$, which indicates a successful optimization, until up to $10$ attempts with different collections of starting values \citep{OpenMx2016package}.} rate of two steps for each condition. Based on our simulation studies, the convergence rate of the proposed two-step model achieved around $90\%$ for all conditions, and the majority of non-convergence cases occurred in the first step. To elaborate, for the conditions with two latent classes, $96$ out of total $288$ conditions reported $100\%$ convergence rate while for the conditions with three latent classes, $12$ out of total $144$ conditions reported $100\%$ convergence rate. Among all conditions with two latent classes, the worst scenario regarding the convergence rate was $121/1121$, indicating that we need to replicate the procedure described in Section \ref{E:step} $1,121$ times to have $1,000$ replications with a convergent solution. Across all scenarios with three latent classes, the worst condition was $134/1134$\footnote{The conditions of these worst cases were the small sample size ($n=500$), unbalanced allocation rate ($1:2$ for two clusters and $1:2:2$ for three latent classes), small residual variance ($1$), small distance between the latent classes (MD was $0.86$), and small difference ($1$) and medium difference ($1.5$) in the knot locations for two or three latent classes, respectively.}.

\subsection{Performance Measures}
\subsubsection*{Performance Measures of the First Part of Simulation Study}
In this section, we evaluated the performance measures of the proposed model across the conditions with fixed knots (i.e., knots without considering variability), under which the proposed model was specified correctly. In the result section, we named the latent classes from left to right as Class 1 (the left cluster) and Class 2 (the right cluster) and called them as Class 1 (the left cluster), Class 2 (the middle cluster) and Class 3 (the right cluster) for the model with two and three pre-specified clusters, respectively. For each parameter of interest, we first calculated each performance metric across $1,000$ replications under each condition with two latent classes and fixed knots. We then summarized each metric across all conditions as the corresponding median and range. 

Tables \ref{tbl:c2_rBias0} and \ref{tbl:c2_empSE0} present the median (range) of the relative bias and empirical SE for each parameter of interest of the two-step model, respectively. We observed that the proposed model generated unbiased point estimates with small empirical SEs when being specified correctly in the first step. Specifically, the magnitude of the relative biases of the growth factor means and growth factor variances across all conditions were under $0.016$ and $0.038$, respectively. In the second step, the median of relative bias of the logistic coefficients was around $-0.010$, although they may be underestimated under conditions with the small sample size (i.e., $n=500$), the small difference in knot locations (i.e., the difference is $1$) and less precise measurements (i.e., $\theta_{\epsilon}=2$). From Table \ref{tbl:c2_empSE0}, the magnitude of empirical SE of all parameters except intercept means and variances were under $0.52$ (i.e., the variances of estimates were under $0.25$), though the median value of empirical SE of $\mu_{\eta_{0}}$ and $\psi_{00}$ were around $0.40$ and $2.50$, respectively. 

\tablehere{3}

\tablehere{4}

Table \ref{tbl:c2_rRMSE0} list the median (range) of relative RMSE of each parameter, which assesses the point estimates holistically. From the table, the model was capable of estimating the parameters accurately in the first step. Under the conditions with two latent classes and fixed knots, the magnitude of the relative RMSEs of the growth factor means and variances were under $0.081$ and $0.296$, respectively. The relative RMSE of the logistic coefficients was relatively larger under some conditions due to their larger relative biases.

\tablehere{5}

Table \ref{tbl:c2_CP0} shows the median (range) of the coverage probability for each parameter of interest of the two-step model with two latent classes under conditions with fixed knots. Overall, the proposed model performed well regarding empirical coverage under the conditions with the relatively large separation between two latent classes and the higher measurement precision. Specifically, coverage probability of all parameters except knots and intercept coefficient $\beta_{0}$ can achieve at least $90\%$ across all conditions with a medium or large separation between the knot locations (i.e., $1.5$ or $2$) and small residual variance (i.e., $\theta_{\epsilon}=1$). 

\tablehere{6}

Additionally, when being specified correctly, the model with three latent classes, similar to that with two clusters, performed well in terms of performance measures, though we noticed that the empirical SE of parameters in the middle cluster were slightly larger than those in the other two groups. 

\subsubsection*{Performance Measures of the Second Part of Simulation Study}
In this section, we assess the robustness of the proposed model by examining the performance measures in the presence of random knots (i.e., the knots with the standard deviation set as $0.3$), under which the model was misspecified. We noted that the relative biases increased slightly and that the empirical SE did not change meaningfully when the proposed model was misspecified, which decreased the performance of relative RMSE and coverage probability. For those conditions under which the model was misspecified, the summary of the relative bias and empirical SE were provided in \ref{Supp:2}.

\subsection{Accuracy}\label{r:clustering}
In this section, we evaluate the accuracy across all conditions we considered in the simulation study. We first calculated mean accuracy across $1,000$ Monte Carlo replications for each condition and plotted these values of the model with two clusters stratified by the MD, the shape of trajectories, and the difference between knot locations in Figure \ref{fig:Accuracy}. Generally, the proposed model grouped trajectories much better under the conditions with larger separations between two latent classes, gauged by the MD and the difference between knot locations. Specifically, the mean accuracy was at least $82\%$ under the conditions with $\text{MD}=1.72$ and $1.5$ or $2.0$ difference between knot locations. Additionally, we noticed that the trajectory shape also affected accuracy. From Figure \ref{fig:Accuracy}, the model performed the best in terms of accuracy when the trajectories from two latent classes were roughly parallel. 

\figurehere{1}

\section{Application}
In this section, we demonstrate how to fit the proposed model to separate nonlinear trajectories and associate the latent classes to the baseline characteristics using the motivating data. We extracted a random subsample ($n=500$) from the Early Childhood Longitudinal Study Kindergarten Cohort: 2010-11 (ECLS-K: 2011) with complete records of repeated mathematics IRT scaled scores, demographic information (sex, race and age in months at each wave), baseline school information (school location and baseline school type), baseline social-economic status (family income and the highest education level between parents), baseline teacher-reported social skills (including interpersonal skills, self-control ability, internalizing problem, externalizing problem), baseline teacher-reported approach to learning, and baseline teacher-reported children behavior question (including inhibitory control and attentional focus)\footnote{The total sample size of ECLS-K: 2011 $n=18174$. The number of entries after removing records with missing values (i.e., rows with any of NaN/-9/-8/-7/-1) is $n=1853$.}. 

ECLS-K: 2011 is a nationally representative longitudinal sample of US children enrolled in about $900$ kindergarten programs beginning with $2010-2011$ school year, where children's mathematics ability was evaluated in nine waves: fall and spring of kindergarten ($2010-2011$), first ($2011-2012$) and second ($2012-2013$) grade, respectively as well as spring of $3^{rd}$ ($2014$), $4^{th}$ ($2015$) and $5^{th}$ ($2016$), respectively. Only about $30\%$ students were assessed in the fall of $2011$ and $2012$ \citep{Le2011ECLS}. In the analysis, we used children's age (in months) rather than their grade-in-school to obtain the time structure with individual measurement occasions. In the subset data, $52\%$ of students were boys, and $48\%$ of students were girls. Additionally, $50\%$ of students were White, $4.8\%$ were Black, $30.4\%$ were Hispanic, $0.2\%$ were Asian, and $14.6\%$ were others. For this analysis, we dichotomized the variable race to be White ($50\%$) and others ($50\%$). At the beginning of the study, $87\%$ and $13\%$ students were from public and private schools, respectively. The covariates including school location (ranged between $1$ and $4$), family income (ranged between $1$ and $18$) and the highest parents' education (ranged between $0$ and $8$) were treated as a continuous variables, and the corresponding mean (SD) was $2.11$ ($1.12$), $11.99$ ($5.34$) and $5.32$ ($1.97$), respectively. 

\subsection*{Step 1}
In the first step, we first fit a latent growth curve model with a linear-linear piecewise functional form and three GMMs with two-, three- and four-class and provided the obtained estimated likelihood, information criteria (AIC and BIC), residual of each latent class in Table \ref{tbl:info}. All four models converged. As mentioned earlier, the BIC is a compelling information criterion when deciding the number of latent classes as it penalizes model complexity and adjusts for sample size \citep{Nylund2007number}. The four fits led to BIC values of $31728.23$, $31531.60$, $31448.99$ and $31478.35$, respectively, which led to the selection of the GMM with three latent classes. 

\tablehere{7}

Table \ref{tbl:est} presents the estimates of growth factors from which we obtained the model implied trajectory of each unobserved group, as shown in Figure \ref{fig:math_curve}. The estimated proportions in Class $1$, $2$ and $3$ were $29.6\%$, $47.8\%$ and $22.6\%$, respectively. On average, students in Class $1$ had the lowest levels of mathematics achievement throughout the entire duration (the fixed effects of the baseline and two slopes were $24.133$, $1.718$ per month, and $0.841$ per month, respectively). On average, students in Class $2$ had a similar initial score and slope for the first stage but relatively lower slope in the second stage (the fixed effects of the baseline and two slopes were $24.498$, $1.730$ per month, and $0.588$ per month, respectively) compared to the students in the Class $1$. Students in Class $3$ had the best mathematics performance on average (the fixed effects of the baseline and two slopes were $36.053$, $2.123$ per month, and $0.605$ per month, respectively). For all three classes, post-knot development in mathematics skills slowed substantially, yet the change to the slower growth rate occurred earlier for Class $1$ and $3$ (around $8$-year old: $91$ and $97$ months, respectively) than Class $2$ (around $9$-year old, $110$ months). Additionally, for each latent class, the estimates of the intercept variance and first slope variance were statistically significant, indicating that each student had a `personal' intercept and pre-knot slope so that a `personal' trajectory of the development in mathematics achievement. 

\tablehere{8}

\figurehere{2}

\subsection*{Step 2}
Table \ref{tbl:step2} summarizes the results of the second step of the GMM to associate latent classes of mathematics achievement trajectories to individual-level covariates. From the table, we noticed that the impacts of some covariates, such as baseline socioeconomic status and teacher-reported skills, may differ with or without other covariates. For example, higher family income, higher parents' education, higher-rated attentional focus and inhibitory control increased the likelihood of being in Class $2$ or Class $3$ in univariable analyses, while these four baseline characteristics only associated with Class $3$ in multivariable analyses. It is reasonable that the effect sizes of the Class $3$ were larger, given its more evident difference from the reference group, as shown in Table \ref{tbl:est} and Figure \ref{fig:math_curve}. However, it is still too rush to conclude that students from families with higher socioeconomic status and/or higher-rated behavior questions were not more likely to be in Class $2$ with all other covariates at the significant level $0.05$ in an exploratory study. Another possible explanation for this phenomenon is multicollinearity.

\tablehere{9}

Figure \ref{fig:correlation} visualizes the correlation matrix of all baseline characteristics, from which we can see that two socioeconomic variables, family income and parents' highest education, were highly correlated ($\rho=0.66$). Additionally, teacher-rated baseline abilities were highly correlated; for example, the correlation of approach to learning with self-control, interpersonal ability, attentional focus, and inhibitory control was $0.68$, $0.72$, $0.79$ and $0.79$, respectively. We then conducted the exploratory factor analysis to address this collinearity issue for socioeconomic variables and teacher-reported abilities. 

The exploratory factor analysis was conducted using the R function \textit{factanal} in the \textit{stats} package \citep{Core2020stat} with $2$ specified factors as suggested by the eigenvalues greater than $1$ (EVG$1$) component retention criterion, scree test \citep{Cattell1966EFA, Cattell1967EFA}, and parallel analysis \citep{Horn1965EFA, Humphreys1969EFA, Humphreys1975EFA}, and `varimax' option to get a type of orthogonal rotation \citep{Kaiser1958EFA}. By using Bartlett's weighted least-squares methods, we obtained the factor scores. Table \ref{tbl:EFA} summarizes the results from the EFA. The first factor differentiates between teacher-rated abilities and teacher-reported problems; the second factor can be interpreted as general socioeconomic status. We then re-ran the second step with the two factors as well as demographic information and school information.

\tablehere{10}

Table \ref{tbl:step2_EFA} summarizes the estimates obtained from the second step with factor scores, demographic information, and school information. From the table, we observed that boys with higher values of the first factor scores, and higher values of the second factor scores were more likely to be in Class $2$ (OR ($95\%$ CI) for sex, factor score $1$ and factor score $2$ was $0.345$ ($0.183$, $0.651$), $1.454$ ($1.090$, $1.939$) and $1.656$ ($1.226$, $2.235$), respectively) or Class $3$ (OR ($95\%$ CI) for sex, factor score $1$ and factor score $2$ was $0.234$ ($0.111$, $0.494$), $2.006$ ($1.408$, $2.858$) and $3.410$ ($2.258$, $5.148$), respectively). It suggests that both socioeconomic variables and teacher-rated abilities were positively associated with mathematics performance while externalizing/internalizing problems were negative associated with mathematics achievement. 

\tablehere{11}

\section{Discussion}
This article extends \citet{Bakk2017two} study to conduct a stepwise analysis for investigating the heterogeneity in nonlinear trajectories. In the first step,  we fit a growth mixture model with a bilinear spline functional form to describe the underlying change pattern of nonlinear trajectories. In the second step, we investigated the associations between the individual-level covariates and the latent classes further. Although this stepwise method follows the recommended approach to fit a FMM model (i.e., group trajectories without including any covariates), it is not our aim to show that this stepwise approach is universally preferred. Based on our understanding, this approach is more suitable for an exploratory study where empirical researchers only have vague assumptions in terms of sample heterogeneity and its possible causes. On the one hand, the two-step model can save the computational budget as we only need to refit the second-step model rather than the whole model when adding or removing covariates. On the other hand, our simulation study showed that the proposed model works well in terms of performance measures and accuracy, especially under mild conditions, such as well-separated latent classes and precise measurements. This stepwise approach can also be utilized to analyze any other types of FMMs in the SEM framework to explore sample heterogeneity. 

\subsection{Methodological Consideration}
Although this stepwise model can expedite the exploratory process, it is still challenging to decide to add which covariates to explain between-class differences, even under the assumption that these covariates only indirectly affect sample heterogeneity. In the psychological and educational domains where the SEM framework is widely used, the candidate pool of covariates is huge, or some variables are highly correlated (i.e., collinearity issue), as shown in the application. 

In the statistical and machine learning (ML) literature, several common approaches to reduce the number of covariates include greedy search, regularization to select covariates based on their corresponding coefficients, principal component analysis (PCA) to transform all features to space with fewer dimensions, and tree-based models (such as regression and classification trees, boosting and bagging). In the SEM framework, the majority of counterparts of the above models have been proposed. For example, \citet{Marcoulides1998search, Marcoulides2003search} proposed to conduct a heuristic specification search algorithm to identify an optimal set of models; \citet{Jacobucci2016regularized, Sun2016regularized, Scharf2019regularized}, demonstrated how to regularize parameters in the SEM framework to reduce the complexity of the model by selecting or removing paths (i.e., variables). Additionally, by applying a tree-based model \citep{Brandmaier2013SEMtree}, \citet{Jacobucci2017SEMtree} captured heterogeneity in trajectories with respect to baseline covariates, where the FMM was compared with the tree-based model in terms of membership components and result interpretation. 

This article proposes to employ the EFA to reduce the dimensions of covariates and address the multicollinearity issue. In this application, we applied the EFA in a process termed as `feature engineering' in the ML literature, where researchers employ the PCA technique to reduce the covariate space and address the multicollinearity issue conventionally, as the interpretation of covariate coefficients is out of the primary interest in the ML literature. In this article, we decided to use the EFA rather than the PCA for two reasons. First, empirical researchers using the SEM framework are more familiar with the EFA as the idea behind it is very similar to another model in the SEM framework, the confirmatory factor analysis (CFA). More importantly, the factors (i.e., latent variables) obtained from the EFA are interpretable so that the estimates obtained from the second step are interpretable, and we then gain valuable insights from an exploratory study. For example, in the application, we concluded that a student with a higher value of the difference between teacher-rated abilities and teacher-reported problems, and/or from a family with higher socioeconomic status was more likely to achieve higher mathematics scores (i.e., in Class $2$ and Class $3$).

Although it is not our aim to comprehensively investigate the EFA, we still want to add two notes about factor retention criteria and factor rotation to empirical researchers. Following \citet{Fabrigar1999EFA}, we used multiple criteria in the application, including the EVG$1$ rule, scree test, and parallel analysis to decide the number of factors; fortunately, all these criteria gave the same decision. \citet{Patil2008EFA} also suggested conducting a subsequent CFA to evaluate the measurement properties of the factors identified by the EFA (if the number of factors is different from multiple criteria). 

Additionally, several analytic rotation techniques have been developed for the EFA with the most fundamental distinction lying in orthogonal and oblique rotation. Orthogonal rotations constrain factors to be uncorrelated, and the procedure, varimax, which we used in the application, is generally regarded as the best one and the most widely used orthogonal rotation in psychological research. One reason for this choice was because of its simplicity and conceptual clarity. More importantly, we assumed that the constructs (i.e., the factor of the socioeconomic variables and that of teacher-rated scores) identified from the set of covariates are independent. However, a considerable amount of theoretical and empirical researchers provided the basis for expecting psychological constructs, such as personality traits, ability, and attitudes, to be associated with each other. Consequently, oblique rotations provide a more realistic and accurate picture of these factors.

One limitation of the proposed two-step model lies in it only allows (generalized) linear models in the second step. If the linear assumption is invalid, we need to resort to other methods, such as structural equation model trees (SEM trees, \citet{Brandmaier2013SEMtree}) or structural equation model forests \citep{Brandmaier2016semForest} to identify the most important covariates by investigating the variables on which the tree splits first \citep{Brandmaier2013SEMtree, Jacobucci2017SEMtree} or the output named `variable importance' \citep{Brandmaier2016semForest}, respectively. Note that \citet{Jacobucci2017SEMtree} pointed out that the interpretations of the FMM and SEM trees are different, and the classes obtained from the SEM tree can be viewed as the clusters of associations between the covariates and trajectories. In other words, the covariates in the SEM trees inform class formation. 

\subsection{Future Research}
One possible future direction of the current study is to build its confirmatory counterpart. Conceptually, the confirmatory model consists of two measurement models and there exists a unidirectional relationship between the factors of the EFA and the latent categorical variable. Additionally, driven by the domain knowledge, the EFA can be replaced with the CFA in the confirmatory model. Additionally, the two-step model is proposed under the assumption that these covariates only have indirect impacts on the sample heterogeneity. It is also possible to develop a model that allows these baseline covariates to simultaneously explain between-group differences and within-group differences by relaxing the assumption. 

\bibliographystyle{apalike}
\bibliography{Paper2}

\appendix
\renewcommand{\theequation}{A.\arabic{equation}}
\setcounter{equation}{0}

\renewcommand{\thesection}{Appendix \Alph{section}}
\renewcommand{\thesubsection}{A.\arabic{subsection}}

\section{{\textbf{Formula Derivation}}}
\subsection{\textbf{The Reparameterizing Procedure for a Fixed Knot}}\label{Supp:1A}
In the original setting of the bilinear spline model, we have three growth factors: an intercept at $t_{0}$ ($\eta_{0}$) and one slope of each stage ($\eta_{1}$ and $\eta_{2}$, respectively). To estimate knots, we may reparameterize the growth factors. For the $i^{th}$ individual, according to Seber and Wild \citep{Seber2003nonlinear}, we may re-expressed them as the measurement at the knot (i.e., $\eta_{0i}+\eta_{1i}\gamma^{(k)}$), the mean of two slopes (i.e., $\frac{\eta_{1i}+\eta_{2i}}{2}$), and the half difference between two slopes (i.e., $\frac{\eta_{2i}-\eta_{1i}}{2}$).

\figurehere{A.1}

\citet{Tishler1981nonlinear} and \citet{Seber2003nonlinear} showed that the regression model with two linear stages can be written as either the minimum or maximum response value of two trajectories. \citet{Liu2019BLSGM} extended such expressions to the latent growth curve modeling framework and showed two forms of bilinear spline for the $i^{th}$ individual in Figure \ref{fig:proj1_2cases}. In the left panel ($\eta_{1i}>\eta_{2i}$), the measurement $y_{ij}$ is always the minimum value of two lines; that is, $y_{ij}=\min{(\eta_{0i}+\eta_{1i}t_{ij}, \eta_{02i}+\eta_{2i}t_{ij})}$. To unify the formula of measurements pre- and post-knot, we express $y_{ij}$ as
\begin{equation}\label{eq:left}
\begin{aligned}
y_{ij} &= \min{(\eta_{0i} + \eta_{1i}t_{ij}, \eta_{02i} + \eta_{2i}t_{ij})}\\
&= \frac{1}{2}\big(\eta_{0i} + \eta_{1i}t_{ij} + \eta_{02i} + \eta_{2i}t_{ij} - 
|\eta_{0i} + \eta_{1i}t_{ij} - \eta_{02i} - \eta_{2i}t_{ij}|\big)\\
&= \frac{1}{2}\big(\eta_{0i} + \eta_{1i}t_{ij} + \eta_{02i} + \eta_{2i}t_{ij}\big) - 
\frac{1}{2}\big(|\eta_{0i} + \eta_{1i}t_{ij} - \eta_{02i} - \eta_{2i}t_{ij}|\big)\\
&= \frac{1}{2}\big(\eta_{0i} + \eta_{02i} + \eta_{1i}t_{ij} + \eta_{2i}t_{ij}\big) - 
\frac{1}{2}\big(\eta_{1i} - \eta_{2i}\big)|t_{ij} - \gamma^{(k)}|\\
&= \eta_{0i}^{'} + \eta_{1i}^{'}\big(t_{ij}-\gamma^{(k)}\big) + \eta_{2i}^{'}|t_{ij} - \gamma^{(k)}|\\
&= \eta_{0i}^{'} + \eta_{1i}^{'}\big(t_{ij}-\gamma^{(k)}\big) + \eta_{2i}^{'}\sqrt{(t_{ij} - \gamma^{(k)})^2},
\end{aligned}
\end{equation}
where $\eta_{0i}^{'}$, $\eta_{1i}^{'}$ and $\eta_{2i}^{'}$ are the measurement at the knot, the mean of two slopes, and the half difference between two slopes. Similarly, the measurement $y_{ij}$ of the bilinear spline in the right panel, in which the measurement $y_{ij}$ is always the maximum value of two lines, has the identical final form in Equation \ref{eq:left}. 

\subsection{\textbf{Class-specific Transformation and Inverse-transformation between Two  Parameter-spaces}}\label{Supp:1B}
Suppose $\boldsymbol{f}: \mathcal{R}^{3}\rightarrow \mathcal{R}^{3}$ is a function, which takes a point $\boldsymbol{\eta}_{i}\in\mathcal{R}^{3}$ as input and produces the vector $\boldsymbol{f}(\boldsymbol{\eta}_{i})\in\mathcal{R}^{3}$ (i.e., $\boldsymbol{\eta}_{i}^{'}\in\mathcal{R}^{3}$) as output. By the multivariate delta method \cite{Lehmann1998Delta}, for an individual in the $k^{th}$ class
\begin{equation}\label{eq:trans_fun}
\boldsymbol{\eta}_{i}^{'}=\boldsymbol{f}(\boldsymbol{\eta}_{i})\sim N\bigg(\boldsymbol{f}(\boldsymbol{\mu_{\eta}}^{[k]}), \boldsymbol{\nabla_{f}}(\boldsymbol{\mu_{\eta}}^{[k]})\boldsymbol{\Psi_{\eta}}^{[k]}\boldsymbol{\nabla}^{T}_{\boldsymbol{f}}(\boldsymbol{\mu_{\eta}}^{[k]})\bigg), 
\end{equation}
where $\boldsymbol{\mu_{\eta}}^{[k]}$ and $\boldsymbol{\Psi_{\eta}}^{[k]}$ are the mean vector and variance-covariance matrix of original class-specific growth factors, respectively, and $\boldsymbol{f}$ is defined as
\begin{equation}\nonumber
\boldsymbol{f}(\boldsymbol{\eta}_{i})=\left(\begin{array}{rrr}
\eta_{0i}+\gamma^{[k]}\eta_{1i} & \frac{\eta_{1i}+\eta_{2i}}{2} & \frac{\eta_{2i}-\eta_{1i}}{2}
\end{array}\right)^{T}.
\end{equation}

Similarly, suppose $\boldsymbol{h}: \mathcal{R}^{3}\rightarrow \mathcal{R}^{3}$ is a function, which takes a point $\boldsymbol{\eta}_{i}^{'}\in\mathcal{R}^{3}$ as input and produces the vector $\boldsymbol{h}(\boldsymbol{\eta}_{i}^{'})\in\mathcal{R}^{3}$ (i.e., $\boldsymbol{\eta}_{i}\in\mathcal{R}^{3}$) as output. By the multivariate delta method,
\begin{equation}\label{eq:inverse_fun}
\boldsymbol{\eta}_{i}=\boldsymbol{h}(\boldsymbol{\eta}_{i}^{'[k]})\sim N\bigg(\boldsymbol{h}(\boldsymbol{\mu}^{'[k]}_{\boldsymbol{\eta}}), \boldsymbol{\nabla_{h}}(\boldsymbol{\mu_{\eta}^{'[k]}})\boldsymbol{\Psi}^{'[k]}_{\boldsymbol{\eta}}\boldsymbol{\nabla}^{T}_{\boldsymbol{h}}(\boldsymbol{\mu_{\eta}^{'[k]}})\bigg), 
\end{equation}
where $\boldsymbol{\mu}^{'[k]}_{\boldsymbol{\eta}}$ and $\boldsymbol{\Psi}^{'[k]}_{\boldsymbol{\eta}}$ are the mean vector and variance-covariance matrix of class-specific reparameterized growth factors, respectively, and $\boldsymbol{h}$ is defined as
\begin{equation}\nonumber
\boldsymbol{h}(\boldsymbol{\eta}_{i}^{'})=\left(\begin{array}{rrr}
\eta^{'}_{0i}-\gamma^{[k]}\eta^{'}_{1i}+\gamma^{[k]}\eta^{'}_{2i} & \eta^{'}_{1i}-\eta^{'}_{2i} & \eta^{'}_{1i}+\eta^{'}_{2i}
\end{array}\right)^{T}.
\end{equation}

Based on Equations (\ref{eq:trans_fun}) and (\ref{eq:inverse_fun}), we can make the transformation between the growth factor means of two parameter-spaces by $\boldsymbol{\mu}^{'[k]}_{\boldsymbol{\eta}}
\approx\boldsymbol{f}(\boldsymbol{\mu}^{[k]}_{\boldsymbol{\eta}})$ and $\boldsymbol{\mu}^{[k]}_{\boldsymbol{\eta}}
\approx\boldsymbol{h}(\boldsymbol{\mu}^{'[k]}_{\boldsymbol{\eta}})$, respectively. We can also define the transformation matrix $\boldsymbol{\nabla_{f}}(\boldsymbol{\mu}^{[k]}_{\boldsymbol{\eta}})$ and $\boldsymbol{\nabla_{h}}(\boldsymbol{\mu_{\eta}^{'[k]}})$ between the variance-covariance matrix of two parameter-spaces as
\begin{equation}\nonumber
\begin{aligned}
\quad\quad\boldsymbol{\Psi}^{'[k]}_{\boldsymbol{\eta}} 
&\approx \boldsymbol{\nabla_{f}}(\boldsymbol{\mu}^{[k]}_{\boldsymbol{\eta}})\boldsymbol{\Psi}^{[k]}_{\boldsymbol{\eta}}\boldsymbol{\nabla}^{T}_{\boldsymbol{f}}(\boldsymbol{\mu}^{[k]}_{\boldsymbol{\eta}})\\
&=\left(\begin{array}{rrr}
1 & \gamma^{[k]} & 0 \\
0 & 0.5 & 0.5 \\
0 & -0.5 & 0.5 
\end{array}\right)\boldsymbol{\Psi}^{[k]}_{\boldsymbol{\eta}}\left(\begin{array}{rrr}
1 & \gamma^{[k]} & 0 \\
0 & 0.5 & 0.5 \\
0 & -0.5 & 0.5
\end{array}\right)^{T} \ \ \ \ \ \ \ \\
\end{aligned}
\end{equation}
and
\begin{equation}\nonumber
\begin{aligned}
\boldsymbol{\Psi_{\eta}}^{[k]} &\approx \boldsymbol{\nabla_{h}}(\boldsymbol{\mu_{\eta}^{'[k]}})\boldsymbol{\Psi}^{'[k]}_{\boldsymbol{\eta}}\boldsymbol{\nabla}^{T}_{\boldsymbol{h}}(\boldsymbol{\mu_{\eta}^{'[k]}})\\
&=\left(\begin{array}{rrr}
1 & -\gamma^{[k]} & \gamma^{[k]} \\0 & 1 & -1 \\0 & 1 & 1 
\end{array}\right) \boldsymbol{\Psi}^{'[k]}_{\boldsymbol{\eta}}\left(\begin{array}{rrrr}
1 & -\gamma^{[k]} & \gamma^{[k]} \\0 & 1 & -1 \\0 & 1 & 1 
\end{array}\right)^{T},
\end{aligned}
\end{equation}
respectively. 

\newpage
\section{\textbf{More Results}}\label{Supp:2}
\tablehere{B.1}
\tablehere{B.2}

\newcolumntype{L}[1]{>{\raggedright\arraybackslash}p{#1}}
\newcolumntype{C}[1]{>{\centering\arraybackslash}p{#1}}
\newcolumntype{R}[1]{>{\raggedleft\arraybackslash}p{#1}}

\renewcommand\thetable{\arabic{table}}
\setcounter{table}{0}

\FloatBarrier
\begin{table}[!ht]
\centering
\begin{threeparttable}
\caption{Performance Metric: Definitions and Estimates}
\begin{tabular}{p{4cm}p{4.5cm}p{5.5cm}}
\hline
\hline
\textbf{Criteria} & \textbf{Definition} & \textbf{Estimate} \\
\hline
Relative Bias & $E_{\hat{\theta}}(\hat{\theta}-\theta)/\theta$ & $\sum_{s=1}^{S}(\hat{\theta}-\theta)/S\theta$ \\
Empirical SE & $\sqrt{Var(\hat{\theta})}$ & $\sqrt{\sum_{s=1}^{S}(\hat{\theta}-\bar{\theta})^{2}/(S-1)}$ \\
Relative RMSE & $\sqrt{E_{\hat{\theta}}(\hat{\theta}-\theta)^{2}}/\theta$ & $\sqrt{\sum_{s=1}^{S}(\hat{\theta}-\theta)^{2}/S}/\theta$ \\
Coverage Probability & $Pr(\hat{\theta}_{\text{low}}\le\theta\le\hat{\theta}_{\text{upper}})$ & $\sum_{s=1}^{S}I(\hat{\theta}_{\text{low},s}\le\theta\le\hat{\theta}_{\text{upper},s})/S$\\
\hline
\hline
\end{tabular}
\label{tbl:metric}
\begin{tablenotes}
\small
\item[1] {$\theta$: the population value of the parameter of interest} \\
\item[2] {$\hat{\theta}$: the estimate of $\theta$} \\
\item[3] {$S$: the number of replications and set as $1,000$ in our simulation study} \\
\item[4] {$s=1,\dots,S$: indexes the replications of the simulation} \\
\item[5] {$\hat{\theta}_{s}$: the estimate of $\theta$ from the $s^{th}$ replication} \\
\item[6] {$\bar{\theta}$: the mean of $\hat{\theta}_{s}$'s across replications} \\
\item[7] {$I()$: an indicator function}
\end{tablenotes}
\end{threeparttable}
\end{table}

\begin{table}[!ht]
\centering
\begin{threeparttable}
\setlength{\tabcolsep}{5pt}
\renewcommand{\arraystretch}{0.6}
\caption{Simulation Design for Two-step GMM with the Bilinear Spline Growth Within-class Model (Unknown Knot) in the Framework of Individual Measurement Occasions}
\begin{tabular}{p{4.0cm} p{6.4cm} p{6.8cm}}
\hline
\hline
\multicolumn{3}{c}{\textbf{Fixed Conditions}} \\
\hline
\textbf{Variables} & \multicolumn{2}{c}{\textbf{Conditions}} \\
\hline
Variance of Intercept & \multicolumn{2}{l}{$\psi_{00}^{(k)}=25$} \\
\hline
Variance of Slopes & \multicolumn{2}{l}{$\psi_{11}^{(k)}=\psi_{22}^{(k)}=1$} \\
\hline
Correlations of GFs & \multicolumn{2}{l}{$\rho^{(k)}=0.3$} \\
\hline
Time (\textit{t}) & \multicolumn{2}{l}{$10$ scaled and equally spaced $t_{j} (j=0, \cdots, J-1, J=10)$} \\
\hline
Individual \textit{t} & \multicolumn{2}{l}{$t_{ij} \sim U(t_{j}-\Delta, t_{j}+\Delta) (j=0, \cdots, J-1; \Delta=0.25)$} \\
\hline
\hline
\multicolumn{3}{c}{\textbf{Manipulated Conditions}} \\
\hline
\hline
\textbf{Variables} & \textbf{Conditions of 2 latent classes} & \textbf{Conditions of 3 latent classes} \\
\hline
Sample Size & $n=500$ or $1000$ & $n=500$ or $1000$ \\
\hline
\multirow{2}{*}{Variance of Knots} & $\psi_{\gamma\gamma}^{(k)}=0.00 (k = 1, 2)$ & $\psi_{\gamma\gamma}^{(k)}=0.00 (k = 1, 2, 3)$ \\
& $\psi_{\gamma\gamma}^{(k)}=0.09 (k = 1, 2)$ & $\psi_{\gamma\gamma}^{(k)}=0.09 (k = 1, 2, 3)$ \\
\hline
\multirow{3}{*}{Ratio of Proportions} & $\pi^{(1)}:\pi^{(2)}=1:1$ & $\pi^{(1)}:\pi^{(2)}:\pi^{(3)}=1:1:1$ \\
& $\pi^{(1)}:\pi^{(2)}=1:2$ & $\pi^{(1)}:\pi^{(2)}:\pi^{(3)}=1:1:2$ \\
& & $\pi^{(1)}:\pi^{(2)}:\pi^{(3)}=1:2:2$ \\
\hline
Residual Variance & $\theta_{\epsilon}^{(k)}=1$ or $2$ & $\theta_{\epsilon}^{(k)}=1$ or $2$ \\
\hline
\multirow{3}{*}{Locations of knots} & $\mu_{\gamma}^{(1)}=4.00$; $\mu_{\gamma}^{(2)}=5.00$ & $\mu_{\gamma}^{(1)}=3.50$; $\mu_{\gamma}^{(2)}=4.50$; $\mu_{\gamma}^{(3)}=5.50$ \\
& $\mu_{\gamma}^{(1)}=3.75$; $\mu_{\gamma}^{(2)}=5.25$ & $\mu_{\gamma}^{(1)}=3.00$; $\mu_{\gamma}^{(2)}=4.50$; $\mu_{\gamma}^{(3)}=6.00$ \\
& $\mu_{\gamma}^{(1)}=3.50$; $\mu_{\gamma}^{(2)}=5.50$ & \\
\hline
Mahalanobis distance (MD) & $d=0.86$ or $1.72$ & $d=0.86$ \\
\hline
\hline
\multicolumn{3}{l}{\textbf{Scenario 1: Different means of initial status and (means of) knot locations}} \\
\hline
\textbf{Variables} & \textbf{Conditions of 2 latent classes} & \textbf{Conditions of 3 latent classes} \\
\hline
Means of Slope 1's & $\mu_{\eta_{1}}^{(k)}=-5$ $(k=1, 2)$ & $\mu_{\eta_{1}}^{(k)}=-5$ $(k=1, 2, 3)$ \\
\hline
Means of Slope 2's
& $\mu_{\eta_{2}}^{(k)}=-2.6$ $(k=1, 2)$ 
& $\mu_{\eta_{2}}^{(k)}=-2.6$ $(k=1, 2, 3)$ \\
\hline
\multirow{2}{*}{Means of Intercepts} & $\mu_{\eta_{0}}^{(1)}=98$, $\mu_{\eta_{0}}^{(2)}=102$ $(d=0.86)$ & $\mu_{\eta_{0}}^{(1)}=96$, $\mu_{\eta_{0}}^{(2)}=100$, $\mu_{\eta_{0}}^{(3)}=104$ \\
& $\mu_{\eta_{0}}^{(1)}=96$, $\mu_{\eta_{0}}^{(2)}=104$ $(d=1.72)$ & \\
\hline
\hline
\multicolumn{3}{l}{\textbf{Scenario 2: Different means of slope 1 and (means of) knot locations}} \\
\hline
\textbf{Variables} & \textbf{Conditions of 2 latent classes} & \textbf{Conditions of 3 latent classes} \\
\hline
Means of Intercepts & $\mu_{\eta_{0}}^{(k)}=100$ $(k=1,2)$ & $\mu_{\eta_{0}}^{(k)}=100$ $(k=1,2,3)$ \\
\hline
Means of Slope 2's & $\mu_{\eta_{2}}^{(k)}=-2$ $(k=1, 2)$ & $\mu_{\eta_{2}}^{(k)}=-2$ $(k=1, 2, 3)$ \\
\hline
\multirow{2}{*}{Means of Slope 1's} & $\mu_{\eta_{1}}^{(1)}=-4.4$, $\mu_{\eta_{1}}^{(2)}=-3.6$ $(d=0.86)$ & $\mu_{\eta_{1}}^{(1)}=-5.2$, $\mu_{\eta_{1}}^{(2)}=-4.4$, $\mu_{\eta_{1}}^{(3)}=-3.6$ \\
& $\mu_{\eta_{1}}^{(1)}=-5.2$, $\mu_{\eta_{1}}^{(2)}=-3.6$ $(d=1.72)$ & \\
\hline
\hline
\multicolumn{3}{l}{\textbf{Scenario 3: Different means of slope 2 and (means of) knot locations}} \\
\hline
\textbf{Variables} & \textbf{Conditions of 2 latent classes} & \textbf{Conditions of 3 latent classes} \\
\hline
Means of Intercepts & $\mu_{\eta_{0}}^{(k)}=100$ $(k=1,2)$ & $\mu_{\eta_{0}}^{(k)}=100$ $(k=1,2,3)$ \\
\hline
Means of Slope 1's & $\mu_{\eta_{1}}^{(k)}=-5$ $(k=1,2)$ & $\mu_{\eta_{1}}^{(k)}=-5$ $(k=1, 2, 3)$ \\
\hline
\multirow{2}{*}{Means of Slope 2's} & $\mu_{\eta_{2}}^{(1)}=-2.6$, $\mu_{\eta_{2}}^{(2)}=-3.4$ $(d=0.86)$ & $\mu_{\eta_{2}}^{(1)}=-1.8$, $\mu_{\eta_{2}}^{(2)}=-2.6$, $\mu_{\eta_{2}}^{(3)}=-3.4$ \\
& $\mu_{\eta_{2}}^{(1)}=-1.8$, $\mu_{\eta_{2}}^{(2)}=-3.4$ $(d=1.72)$ & \\
\hline
\hline
\end{tabular}
\label{tbl:simu}
\end{threeparttable}
\end{table}

\begin{table}[!ht]
\centering
\begin{threeparttable}
\setlength{\tabcolsep}{5pt}
\renewcommand{\arraystretch}{0.6}
\caption{Median (Range) of the Relative Bias over $1,000$ Replications of Parameters of Interest under the Conditions with Fixed Knots and 2 Latent Classes}
\begin{tabular}{p{3cm}p{2cm}R{5cm}R{5cm}}
\hline
\hline
& \textbf{Para.} & \textbf{Latent Class $1$} & \textbf{Latent Class $2$} \\
\hline
\hline
\multirow{4}{*}{\textbf{Mean}} 
& $\mu_{\eta_{0}}$ & $0.000$ ($0.000$, $0.001$) & $0.000$ ($-0.001$, $0.000$) \\
& $\mu_{\eta_{1}}$ & $0.000$ ($-0.008$, $0.003$) & $0.001$ ($-0.001$, $0.012$) \\
& $\mu_{\eta_{2}}$ & $0.000$ ($-0.009$, $0.016$) & $-0.002$ ($-0.012$, $0.003$) \\
& $\mu_{\gamma}$ & $0.000$ ($-0.001$, $0.002$) & $0.000$ ($-0.001$, $0.002$) \\
\hline
\hline
\multirow{4}{*}{\textbf{Variance}} 
& $\psi_{00}$ & $-0.002$ ($-0.014$, $0.006$) & $-0.005$ ($-0.031$, $0.005$) \\
& $\psi_{11}$ & $-0.005$ ($-0.028$, $0.028$) & $-0.007$ ($-0.038$, $0.003$) \\
& $\psi_{22}$ & $-0.005$ ($-0.026$, $0.031$) & $-0.007$ ($-0.037$, $0.005$) \\
\hline
\hline
\multirow{4}{*}{\textbf{Path Coef.}} 
& $\beta_{0}$ & ---\tnote{1} & $-0.009$ (NA\tnote{2}, NA) \\
& $\beta_{1}$ & --- & $-0.012$ ($-0.225$, $0.018$) \\
& $\beta_{2}$ & --- & $-0.010$ ($-0.218$, $0.015$) \\
\hline
\hline
\end{tabular}
\label{tbl:c2_rBias0}
\begin{tablenotes}
\small
\item[1] ---: when fitting the proposed model, we set the first latent class as the reference group; accordingly, the coefficients of that class do not exist.
\item[2] NA: Note that for the conditions with balanced allocation, the population value of $\beta_{0}=0$ and its relative bias goes infinity. The bias median (range) of $\beta_{0}$ is $-0.002$ ($-0.070$, $0.017$).  
\end{tablenotes}
\end{threeparttable}
\end{table}

\begin{table}[!ht]
\centering
\begin{threeparttable}
\setlength{\tabcolsep}{5pt}
\renewcommand{\arraystretch}{0.6}
\caption{Median (Range) of the Empirical SE over $1,000$ Replications of Parameters of Interest under the Conditions with Fixed Knots and 2 Latent Classes}
\begin{tabular}{p{3cm}p{2cm}R{5cm}R{5cm}}
\hline
\hline
& \textbf{Para.} & \textbf{Latent Class $1$} & \textbf{Latent Class $2$} \\
\hline
\hline
\multirow{4}{*}{\textbf{Mean}} 
& $\mu_{\eta_{0}}$ & $0.422$ ($0.242$, $0.933$) & $0.336$ ($0.198$, $0.709$) \\
& $\mu_{\eta_{1}}$ & $0.101$ ($0.051$, $0.276$) & $0.073$ ($0.042$, $0.175$) \\
& $\mu_{\eta_{2}}$ & $0.100$ ($0.054$, $0.276$) & $0.072$ ($0.042$, $0.160$) \\
& $\mu_{\gamma}$ & $0.039$ ($0.017$, $0.110$) & $0.046$ ($0.020$, $0.134$) \\
\hline
\hline
\multirow{4}{*}{\textbf{Variance}} 
& $\psi_{00}$ & $2.662$ ($1.692$, $5.073$) & $2.173$ ($1.423$, $3.942$) \\
& $\psi_{11}$ & $0.124$ ($0.073$, $0.296$) & $0.093$ ($0.059$, $0.168$) \\
& $\psi_{22}$ & $0.126$ ($0.072$, $0.286$) & $0.095$ ($0.062$, $0.178$) \\
\hline
\hline
\multirow{4}{*}{\textbf{Path Coef.}} 
& $\beta_{0}$ & ---\tnote{1} & $0.168$ ($0.083$, $0.516$) \\
& $\beta_{1}$ & --- & $0.120$ ($0.080$, $0.200$) \\
& $\beta_{2}$ & --- & $0.124$ ($0.082$, $0.198$) \\
\hline
\hline
\end{tabular}
\label{tbl:c2_empSE0}
\begin{tablenotes}
\small
\item[1] ---: when fitting the proposed model, we set the first latent class as the reference group; accordingly, the coefficients of that class do not exist.
\end{tablenotes}
\end{threeparttable}
\end{table}

\begin{table}[!ht]
\centering
\begin{threeparttable}
\setlength{\tabcolsep}{5pt}
\renewcommand{\arraystretch}{0.6}
\caption{Median (Range) of the Relative RMSE over $1,000$ Replications of Parameters of Interest under the Conditions with Fixed Knots and 2 Latent Classes}
\begin{tabular}{p{3cm}p{2cm}R{5cm}R{5cm}}
\hline
\hline
& \textbf{Para.} & \textbf{Latent Class $1$} & \textbf{Latent Class $2$} \\
\hline
\hline
\multirow{4}{*}{\textbf{Mean}} 
& $\mu_{\eta_{0}}$ & $0.004$ ($0.002$, $0.009$) & $0.003$ ($0.002$, $0.007$) \\
& $\mu_{\eta_{1}}$ & $-0.021$ ($-0.063$, $-0.010$) & $-0.016$ ($-0.045$, $-0.009$) \\
& $\mu_{\eta_{2}}$ & $-0.045$ ($-0.112$, $-0.020$) & $-0.028$ ($-0.081$, $-0.012$) \\
& $\mu_{\gamma}$ & $0.010$ ($0.005$, $0.028$) & $0.009$ ($0.004$, $0.027$) \\
\hline
\hline
\multirow{4}{*}{\textbf{Variance}} 
& $\psi_{00}$ & $0.106$ ($0.068$, $0.203$) & $0.087$ ($0.057$, $0.161$) \\
& $\psi_{11}$ & $0.124$ ($0.074$, $0.296$) & $0.093$ ($0.060$, $0.172$) \\
& $\psi_{22}$ & $0.126$ ($0.072$, $0.288$) & $0.095$ ($0.062$, $0.182$) \\
\hline
\hline
\multirow{4}{*}{\textbf{Path Coef.}} 
& $\beta_{0}$ & ---\tnote{1} & NA\tnote{2} ($0.121$, NA) \\
& $\beta_{1}$ & --- & $0.297$ ($0.197$, $0.542$) \\
& $\beta_{2}$ & --- & $0.234$ ($0.155$, $0.431$) \\
\hline
\hline
\end{tabular}
\label{tbl:c2_rRMSE0}
\begin{tablenotes}
\small
\item[1] ---: when fitting the proposed model, we set the first latent class as the reference group; accordingly, the coefficients of that class do not exist.
\item[2] NA: Note that for the conditions with balanced allocation, the population value of $\beta_{0}=0$ and its relative RMSE goes infinity. The RMSE median (range) of $\beta_{0}$ is $0.168$ ($0.083$, $0.521$).  
\end{tablenotes}
\end{threeparttable}
\end{table}

\begin{table}[!ht]
\centering
\begin{threeparttable}
\setlength{\tabcolsep}{5pt}
\renewcommand{\arraystretch}{0.6}
\caption{Median (Range) of the Coverage Probabilities over $1,000$ Replications of Parameters of Interest under the Conditions with Fixed Knots and 2 Latent Classes}
\begin{tabular}{p{2.0cm}p{1cm}rrrr}
\hline
\hline
\multicolumn{6}{c}{\textbf{Small Separation between the Knots Locations}} \\
\hline
\hline
& \textbf{Para.} & \multicolumn{2}{c}{\textbf{Latent Class $1$}} & \multicolumn{2}{c}{\textbf{Latent Class $2$}} \\
\hline
& & Small Residuals & Large Residuals & Small Residuals & Large Residuals \\
\hline
\hline
\multirow{4}{*}{\textbf{Mean}} 
& $\mu_{\eta_{0}}$ & $0.937$ ($0.913$, $0.961$) & $0.915$ ($0.866$, $0.950$) & $0.942$ ($0.920$, $0.971$) & $0.919$ ($0.867$, $0.952$) \\
& $\mu_{\eta_{1}}$ & $0.919$ ($0.861$, $0.948$) & $0.874$ ($0.766$, $0.942$) & $0.936$ ($0.901$, $0.962$) & $0.904$ ($0.819$, $0.941$) \\
& $\mu_{\eta_{2}}$ & $0.926$ ($0.849$, $0.949$) & $0.893$ ($0.747$, $0.940$) & $0.938$ ($0.888$, $0.956$) & $0.913$ ($0.855$, $0.949$) \\
& $\mu_{\gamma}$ & $0.629$ ($0.493$, $0.724$) & $0.476$ ($0.290$, $0.623$) & $0.522$ ($0.406$, $0.685$) & $0.355$ ($0.227$, $0.541$) \\
\hline
\hline
\multirow{4}{*}{\textbf{Variance}} 
& $\psi_{00}$ & $0.939$ ($0.916$, $0.954$) & $0.932$ ($0.896$, $0.950$) & $0.939$ ($0.927$, $0.957$) & $0.925$ ($0.888$, $0.963$) \\
& $\psi_{11}$ & $0.933$ ($0.878$, $0.950$) & $0.921$ ($0.831$, $0.957$) & $0.935$ ($0.911$, $0.966$) & $0.927$ ($0.877$, $0.947$) \\
& $\psi_{22}$ & $0.929$ ($0.862$, $0.950$) & $0.904$ ($0.809$, $0.935$) & $0.938$ ($0.902$, $0.961$) & $0.930$ ($0.888$, $0.957$) \\
\hline
\hline
\multirow{4}{*}{\textbf{Path Coef.}} 
& $\beta_{0}$ & ---\tnote{1} & --- & $0.789$ ($0.665$, $0.854$) & $0.643$ ($0.502$, $0.739$) \\
& $\beta_{1}$ & --- & --- & $0.950$ ($0.935$, $0.960$) & $0.936$ ($0.891$, $0.957$) \\
& $\beta_{2}$ & --- & --- & $0.944$ ($0.930$, $0.959$) & $0.933$ ($0.873$, $0.958$) \\
\hline
\hline
\multicolumn{6}{c}{\textbf{Medium Separation between the Knots Locations}} \\
\hline
\hline
& \textbf{Para.} & \multicolumn{2}{c}{\textbf{Latent Class $1$}} & \multicolumn{2}{c}{\textbf{Latent Class $2$}} \\
\hline
& & Small Residuals & Large Residuals & Small Residuals & Large Residuals \\
\hline
\hline
\multirow{4}{*}{\textbf{Mean}} 
& $\mu_{\eta_{0}}$ & $0.944$ ($0.918$, $0.959$) & $0.929$ ($0.899$, $0.951$) & $0.943$ ($0.923$, $0.957$) & $0.932$ ($0.905$, $0.955$) \\
& $\mu_{\eta_{1}}$ & $0.938$ ($0.897$, $0.957$) & $0.922$ ($0.833$, $0.951$) & $0.947$ ($0.917$, $0.959$) & $0.932$ ($0.884$, $0.959$) \\
& $\mu_{\eta_{2}}$ & $0.935$ ($0.910$, $0.948$) & $0.913$ ($0.835$, $0.947$) & $0.940$ ($0.913$, $0.959$) & $0.934$ ($0.883$, $0.954$) \\
& $\mu_{\gamma}$ & $0.814$ ($0.786$, $0.854$) & $0.740$ ($0.684$, $0.800$) & $0.767$ ($0.721$, $0.833$) & $0.682$ ($0.626$, $0.780$) \\
\hline
\hline
\multirow{4}{*}{\textbf{Variance}} 
& $\psi_{00}$ & $0.940$ ($0.925$, $0.953$) & $0.935$ ($0.912$, $0.955$) & $0.944$ ($0.927$, $0.953$) & $0.939$ ($0.901$, $0.950$) \\
& $\psi_{11}$ & $0.939$ ($0.905$, $0.952$) & $0.929$ ($0.853$, $0.953$) & $0.939$ ($0.914$, $0.961$) & $0.937$ ($0.909$, $0.952$) \\
& $\psi_{22}$ & $0.930$ ($0.906$, $0.958$) & $0.920$ ($0.878$, $0.951$) & $0.939$ ($0.917$, $0.962$) & $0.934$ ($0.889$, $0.951$) \\
\hline
\hline
\multirow{4}{*}{\textbf{Path Coef.}} 
& $\beta_{0}$ & --- & --- & $0.858$ ($0.782$, $0.905$) & $0.770$ ($0.658$, $0.839$) \\
& $\beta_{1}$ & --- & --- & $0.954$ ($0.937$, $0.961$) & $0.944$ ($0.921$, $0.965$) \\
& $\beta_{2}$ & --- & --- & $0.949$ ($0.934$, $0.964$) & $0.942$ ($0.923$, $0.961$) \\
\hline
\hline
\multicolumn{6}{c}{\textbf{Large Separation between the Knots Locations}} \\
\hline
\hline
& \textbf{Para.} & \multicolumn{2}{c}{\textbf{Latent Class $1$}} & \multicolumn{2}{c}{\textbf{Latent Class $2$}} \\
\hline
& & Small Residuals & Large Residuals & Small Residuals & Large Residuals \\
\hline
\hline
\multirow{4}{*}{\textbf{Mean}} 
& $\mu_{\eta_{0}}$ & $0.946$ ($0.931$, $0.955$) & $0.938$ ($0.921$, $0.965$) & $0.946$ ($0.932$, $0.959$) & $0.940$ ($0.921$, $0.967$) \\
& $\mu_{\eta_{1}}$ & $0.938$ ($0.921$, $0.959$) & $0.936$ ($0.875$, $0.953$) & $0.947$ ($0.926$, $0.958$) & $0.937$ ($0.893$, $0.961$) \\
& $\mu_{\eta_{2}}$ & $0.939$ ($0.907$, $0.956$) & $0.928$ ($0.876$, $0.951$) & $0.949$ ($0.937$, $0.964$) & $0.940$ ($0.916$, $0.955$) \\
& $\mu_{\gamma}$ & $0.952$ ($0.935$, $0.970$) & $0.946$ ($0.935$, $0.961$) & $0.950$ ($0.933$, $0.965$) & $0.946$ ($0.932$, $0.960$) \\
\hline
\hline
\multirow{4}{*}{\textbf{Variance}} 
& $\psi_{00}$ & $0.946$ ($0.929$, $0.957$) & $0.944$ ($0.916$, $0.963$) & $0.943$ ($0.916$, $0.958$) & $0.942$ ($0.921$, $0.959$) \\
& $\psi_{11}$ & $0.938$ ($0.917$, $0.952$) & $0.934$ ($0.859$, $0.951$) & $0.942$ ($0.918$, $0.955$) & $0.938$ ($0.902$, $0.956$) \\
& $\psi_{22}$ & $0.935$ ($0.910$, $0.950$) & $0.928$ ($0.857$, $0.951$) & $0.946$ ($0.925$, $0.959$) & $0.938$ ($0.919$, $0.953$) \\
\hline
\hline
\multirow{4}{*}{\textbf{Path Coef.}} 
& $\beta_{0}$ & --- & --- & $0.892$ ($0.825$, $0.924$) & $0.805$ ($0.703$, $0.865$) \\
& $\beta_{1}$ & --- & --- & $0.950$ ($0.927$, $0.958$) & $0.949$ ($0.937$, $0.960$) \\
& $\beta_{2}$ & --- & --- & $0.950$ ($0.934$, $0.964$) & $0.946$ ($0.924$, $0.958$) \\
\hline
\hline
\end{tabular}
\label{tbl:c2_CP0}
\begin{tablenotes}
\small
\item[1]---: when fitting the proposed model, we set the first latent class as the reference group; accordingly, the coefficients of that class do not exist.
\end{tablenotes}
\end{threeparttable}
\end{table}

\begin{table}[!ht]
\centering
\begin{threeparttable}
\setlength{\tabcolsep}{5pt}
\renewcommand{\arraystretch}{0.6}
\caption{Summary of Model Fit Information For the Bilinear Spline Models}
\begin{tabular}{lrrrrrrr}
\hline
\hline
\multicolumn{8}{c}{\textbf{Bilinear Spline Growth Mixture Model with Different \# of Latent Classes}} \\
\hline
\textbf{Model} & \textbf{-2LL} & \textbf{AIC} & \textbf{BIC} & \textbf{Residual $1$} & \textbf{Residual $2$} & \textbf{Residual $3$} & \textbf{Residual $4$} \\
\hline
$1$-Class & $31659.87$ & $31681.87$& $31728.23$ & $35.60$ & $-$ & $-$ & $-$\\
$2$-Class & $31388.67$ & $31434.67$& $31531.60$ & $28.57$ & $35.02$ & $-$ & $-$ \\
$3$-Class & $31231.48$ & $31301.48$& $31448.99$ & $28.47$ & $33.89$ & $32.03$ & $-$ \\
$4$-Class & $31186.26$ & $31280.26$& $31478.35$ & $26.78$ & $32.51$ & $33.36$ & $26.63$ \\
\hline
\hline
\end{tabular}
\label{tbl:info}
\begin{tablenotes}
\small
\item[1] $-$ indicates that the metric was not available for the model.
\end{tablenotes}
\end{threeparttable}
\end{table}

\begin{table}[!ht]
\centering
\begin{threeparttable}
\setlength{\tabcolsep}{5pt}
\renewcommand{\arraystretch}{0.6}
\caption{Estimates of Bilinear Spline Growth Mixture Model with $3$ Latent Classes}
\begin{tabular}{lrrrrrr}
\hline
\hline
& \multicolumn{2}{c}{\textbf{Class 1}} & \multicolumn{2}{c}{\textbf{Class 2}}& \multicolumn{2}{c}{\textbf{Class 3}} \\
\hline
\textbf{Mean} & Estimate (SE) & P value & Estimate (SE) & P value & Estimate (SE) & P value \\
\hline
\textbf{Intercept$^{1}$} & $24.133$ ($1.250$) & $<0.0001^{\ast}$ & $24.498$ ($0.813$) & $<0.0001^{\ast}$ & $36.053$ ($1.729$) & $<0.0001^{\ast}$ \\
\textbf{Slope $1$} & $1.718$ ($0.052$) & $<0.0001^{\ast}$ & $1.730$ ($0.024$) & $<0.0001^{\ast}$ & $2.123$ ($0.035$) & $<0.0001^{\ast}$ \\
\textbf{Slope $2$} & $0.841$ ($0.031$) & $<0.0001^{\ast}$ & $0.588$ ($0.032$) & $<0.0001^{\ast}$ & $0.605$ ($0.027$) & $<0.0001^{\ast}$ \\
\textbf{Knot} & $90.788$ ($0.733$) & $<0.0001^{\ast}$ & $109.653$ ($0.634$) & $<0.0001^{\ast}$ & $97.610$ ($0.068$) & $<0.0001^{\ast}$ \\
\hline
\hline
\textbf{Variance} & Estimate (SE) & P value & Estimate (SE) & P value & Estimate (SE) & P value \\
\hline
\textbf{Intercept} & $79.696$ ($17.419$) & $<0.0001^{\ast}$ & $77.302$ ($11.973$) & $<0.0001^{\ast}$ & $211.198$ ($36.057$) & $<0.0001^{\ast}$ \\
\textbf{Slope $1$} & $0.104$ ($0.023$) & $<0.0001^{\ast}$ & $0.026$ ($0.007$) & $0.0002^{\ast}$ & $0.065$ ($0.017$) & $0.0001^{\ast}$ \\
\textbf{Slope $2$} & $0.049$ ($0.011$) & $<0.0001^{\ast}$ & $0.012$ ($0.011$) & $0.2753$ & $-0.002$ ($0.006$) & $0.7389$ \\
\hline
\hline
\end{tabular}
\label{tbl:est}
\begin{tablenotes}
\small
\item[1] Intercept was defined as mathematics IRT scores at 60-month old in this case.
\item[2] $^{\ast}$ indicates statistical significance at $0.05$ level.
\end{tablenotes}
\end{threeparttable}
\end{table}

\begin{table}
\centering
\begin{threeparttable}
\setlength{\tabcolsep}{5pt}
\renewcommand{\arraystretch}{0.6}
\caption{Odds Ratio (OR) \& $95\%$ Confidence Interval (CI) of Individual-level Predictor of Latent Class in Mathematics Achievement(Reference group: Class $1$)}
\begin{tabular}{p{6.8cm}rrrr}
\hline
\hline
& \multicolumn{4}{c}{\textbf{Class $2$}} \\
\hline
\textbf{Predictor} & \multicolumn{2}{c}{\textbf{Uni-variable}} & \multicolumn{2}{c}{\textbf{Multi-variable}} \\
\hline
& OR & $95\%$ CI & OR & $95\%$ CI \\
\hline
\textbf{Sex}($0-$Boy; $1-$Girl) & $0.435$ & ($0.254$, $0.745$)$^{\ast}$ & $0.332$ & ($0.174$, $0.633$)$^{\ast}$ \\
\hline
\textbf{Race}($0-$White; $1-$Other) & $0.764$ & ($0.455$, $1.281$) & $1.249$ & ($0.624$, $2.498$) \\
\hline
\textbf{School Location} & $1.407$ & ($1.093$, $1.811$)$^{\ast}$ & $1.357$ & ($0.981$, $1.877$) \\
\hline
\textbf{Parents' Highest Education} & $1.208$ & ($1.051$, $1.388$)$^{\ast}$ & $1.155$ & ($0.933$, $1.431$) \\
\hline
\textbf{Income} & $1.074$ & ($1.023$, $1.128$)$^{\ast}$ & $1.067$ & ($0.987$, $1.154$) \\
\hline
\textbf{School Type}($0-$Public; $1-$Private) & $0.573$ & ($0.250$, $1.317$) & $0.442$ & ($0.149$, $1.313$) \\
\hline
\textbf{Approach to Learning} & $1.305$ & ($0.883$, $1.929$) & $0.957$ & ($0.384$, $2.389$) \\
\hline
\textbf{Self-control} & $1.146$ & ($0.764$, $1.718$) & $0.663$ & ($0.272$, $1.616$) \\
\hline
\textbf{Interpersonal Skills} & $1.479$ & ($0.959$, $2.282$) & $1.276$ & ($0.513$, $3.175$) \\
\hline 
\textbf{External Prob Behavior} & $0.858$ & ($0.559$, $1.319$) & $1.391$ & ($0.571$, $3.386$) \\
\hline
\textbf{Internal Prob Behavior} & $1.139$ & ($0.658$, $1.972$) & $1.190$ & ($0.589$, $2.406$) \\
\hline
\textbf{Attentional Focus} & $1.251$ & ($1.035$, $1.511$)$^{\ast}$ & $1.139$ & ($0.764$, $1.698$) \\
\hline
\textbf{Inhibitory Control} & $1.238$ & ($1.007$, $1.520$)$^{\ast}$ & $1.557$ & ($0.915$, $2.649$) \\
\hline
\hline
& \multicolumn{4}{c}{\textbf{Class $3$}} \\
\hline
\textbf{Predictor} & \multicolumn{2}{c}{\textbf{Uni-variable}} & \multicolumn{2}{c}{\textbf{Multi-variable}} \\
\hline
& OR & $95\%$ CI & OR & $95\%$ CI \\
\hline
\textbf{Sex}($0-$Boy; $1-$Girl) & $0.379$ & ($0.205$, $0.700$)$^{\ast}$ & $0.212$ & ($0.098$, $0.459$)$^{\ast}$ \\
\hline
\textbf{Race}($0-$White; $1-$Other) & $0.397$ & ($0.219$, $0.721$)$^{\ast}$ & $0.943$ & ($0.429$, $2.073$) \\
\hline
\textbf{School Location} & $1.266$ & ($0.957$, $1.676$) & $1.211$ & ($0.835$, $1.755$) \\
\hline
\textbf{Parents' Highest Education} & $1.713$ & ($1.418$, $2.068$)$^{\ast}$ & $1.345$ & ($1.043$, $1.734$)$^{\ast}$ \\
\hline
\textbf{Income} & $1.241$ & ($1.155$, $1.334$)$^{\ast}$ & $1.195$ & ($1.083$, $1.318$)$^{\ast}$ \\
\hline
\textbf{School Type}($0-$Public; $1-$Private) & $1.437$ & ($0.661$, $3.124$) & $0.665$ & ($0.234$, $1.892$) \\
\hline
\textbf{Approach to Learning} & $2.624$ & ($1.590$, $4.332$)$^{\ast}$ & $5.363$ & ($1.731$, $16.612$)$^{\ast}$ \\
\hline
\textbf{Self-control} & $1.436$ & ($0.903$, $2.284$) & $0.414$ & ($0.136$, $1.265$) \\
\hline
\textbf{Interpersonal Skills} & $1.740$ & ($1.057$, $2.862$)$^{\ast}$ & $0.771$ & ($0.269$, $2.209$) \\
\hline 
\textbf{External Prob Behavior} & $0.761$ & ($0.451$, $1.283$) & $1.565$ & ($0.561$, $4.367$) \\
\hline
\textbf{Internal Prob Behavior} & $0.787$ & ($0.405$, $1.532$) & $1.170$ & ($0.488$, $2.808$) \\
\hline
\textbf{Attentional Focus} & $1.601$ & ($1.253$, $2.045$)$^{\ast}$ & $1.095$ & ($0.671$, $1.787$)$^{\ast}$ \\
\hline
\textbf{Inhibitory Control} & $1.439$ & ($1.116$, $1.855$)$^{\ast}$ & $1.324$ & ($0.720$, $2.434$)$^{\ast}$ \\
\hline
\hline
\end{tabular}
\label{tbl:step2}
\begin{tablenotes}
\small
\item[1] $^{\ast}$ indicates $95\%$ confidence interval excluded $1$.
\end{tablenotes}
\end{threeparttable}
\end{table}

\begin{table}[!ht]
\centering
\begin{threeparttable}
\caption{Exploratory Factor Analysis of Socioeconomic Variables and Teacher-reported Abilities}
\begin{tabular}{lrr}
\hline
\hline
\multicolumn{3}{c}{\textbf{Factor Loadings}} \\
\hline
\textbf{Baseline Characteristics} & \textbf{Factor 1} & \textbf{Factor 2} \\
\hline
\textbf{Parents' Highest Education} & $0.10$ & $0.76$ \\
\hline
\textbf{Family Income} & $0.03$ & $0.86$ \\  
\hline
\textbf{Approach to Learning} & $0.90$ & $0.04$ \\ 
\hline
\textbf{Self-control} & $0.77$ & $0.08$ \\  
\hline
\textbf{Interpersonal Skills} & $0.76$ & $0.05$ \\
\hline 
\textbf{External Prob Behavior} & $-0.72$ & $0.00$ \\ 
\hline
\textbf{Internal Prob Behavior} & $-0.24$ & $-0.07$ \\
\hline
\textbf{Attentional Focus} & $0.83$ & $0.07$ \\
\hline
\textbf{Inhibitory Control} & $0.89$ & $0.01$ \\
\hline
\hline
\multicolumn{3}{c}{\textbf{Explained Variance}} \\
\hline
 & \textbf{Factor 1} & \textbf{Factor 2} \\
\hline
\textbf{SS Loadings} & $4.04$ & $1.34$ \\
\hline
\textbf{Proportion Variance} & $0.45$ & $0.15$ \\
\hline
\textbf{Cumulative Variance} & $0.45$ & $0.60$ \\
\hline
\hline
\end{tabular}
\label{tbl:EFA}
\end{threeparttable}
\end{table}

\begin{table}
\centering
\begin{threeparttable}
\setlength{\tabcolsep}{5pt}
\renewcommand{\arraystretch}{0.6}
\caption{Odds Ratio (OR) \& $95\%$ Confidence Interval (CI) of Factor Scores, Demographic Information and School Information of Latent Class in Mathematics Achievement (Reference group: Class $1$)}
\begin{tabular}{p{6.8cm}rrrr}
\hline
\hline
\textbf{Predictor} & \multicolumn{2}{c}{\textbf{Class $2$}} & \multicolumn{2}{c}{\textbf{Class $3$}} \\
\hline
& OR & $95\%$ CI & OR & $95\%$ CI \\
\hline
\textbf{Sex}($0-$Boy; $1-$Girl) & $0.345$ & ($0.183$, $0.651$)$^{\ast}$ & $0.234$ & ($0.111$, $0.494$)$^{\ast}$ \\
\hline
\textbf{Race}($0-$White; $1-$Other) & $1.221$ & ($0.638$, $2.339$) & $1.021$ & ($0.486$, $2.145$) \\
\hline
\textbf{School Type}($0-$Public; $1-$Private) & $0.439$ & ($0.149$, $1.291$) & $0.709$ & ($0.244$, $2.056$) \\
\hline
\textbf{School Location} & $1.333$ & ($0.995$, $1.786$) & $1.133$ & ($0.806$, $1.593$) \\
\hline
\textbf{Factor $1$} & $1.454$ & ($1.090$, $1.939$)$^{\ast}$ & $2.006$ & ($1.408$, $2.858$)$^{\ast}$ \\
\hline
\textbf{Factor $2$} & $1.656$ & ($1.226$, $2.235$)$^{\ast}$ & $3.410$ & ($2.258$, $5.148$)$^{\ast}$ \\
\hline
\hline
\end{tabular}
\label{tbl:step2_EFA}
\begin{tablenotes}
\small
\item[1] $^{\ast}$ indicates $95\%$ confidence interval excluded $1$.
\end{tablenotes}
\end{threeparttable}
\end{table}

\renewcommand\thetable{B.\arabic{table}}
\setcounter{table}{0}
\begin{table}[!ht]
\centering
\begin{threeparttable}
\setlength{\tabcolsep}{5pt}
\renewcommand{\arraystretch}{0.6}
\caption{Median (Range) of the Relative Bias over $1,000$ Replications of Parameters of Interest under the Conditions with Random Knots of the Standard Deviation of $0.3$ and 2 Latent Classes}
\begin{tabular}{p{3cm}p{2cm}R{5cm}R{5cm}}
\hline
\hline
& \textbf{Para.} & \textbf{Latent Class $1$} & \textbf{Latent Class $2$} \\
\hline
\hline
\multirow{4}{*}{\textbf{Mean}} 
& $\mu_{\eta_{0}}$ & $-0.003$ ($-0.009$, $0.003$) & $0.002$ ($0.000$, $0.007$) \\
& $\mu_{\eta_{1}}$ & $0.008$ ($-0.009$, $0.029$) & $-0.009$ ($-0.024$, $0.007$) \\
& $\mu_{\eta_{2}}$ & $0.033$ ($0.007$, $0.098$) & $-0.019$ ($-0.060$, $0.001$) \\
& $\mu_{\gamma}$ & $-0.005$ ($-0.016$, $0.004$) & $0.003$ ($-0.005$, $0.013$) \\
\hline
\hline
\multirow{4}{*}{\textbf{Variance}} 
& $\psi_{00}$ & $-0.001$ ($-0.069$, $0.037$) & $-0.016$ ($-0.055$, $0.006$) \\
& $\psi_{11}$ & $-0.076$ ($-0.126$, $-0.040$) & $-0.030$ ($-0.083$, $-0.008$) \\
& $\psi_{22}$ & $-0.015$ ($-0.061$, $0.137$) & $-0.057$ ($-0.089$, $0.179$) \\
\hline
\hline
\multirow{4}{*}{\textbf{Path Coef.}} 
& $\beta_{0}$ & ---\tnote{1} & $-0.055$ (NA\tnote{2}, NA) \\
& $\beta_{1}$ & --- & $-0.042$ ($-0.332$, $0.013$) \\
& $\beta_{2}$ & --- & $-0.038$ ($-0.332$, $0.019$) \\
\hline
\hline
\end{tabular}
\label{tbl:c2_Bias3}
\begin{tablenotes}
\small
\item[1] ---: when fitting the proposed model, we set the first latent class as the reference group; accordingly, the coefficients of that class do not exist.
\item[2] NA: Note that for the conditions with balanced allocation, the population value of $\beta_{0}=0$ and its relative bias goes infinity. The bias median (range) of $\beta_{0}$ is $-0.015$ ($-0.204$, $0.118$).  
\end{tablenotes}
\end{threeparttable}
\end{table}

\begin{table}[!ht]
\centering
\begin{threeparttable}
\setlength{\tabcolsep}{5pt}
\renewcommand{\arraystretch}{0.6}
\caption{Median (Range) of the Empirical SE over $1,000$ Replications of Parameters of Interest under the Conditions with Random Knots of the Standard Deviation of $0.3$ and 2 Latent Classes}
\begin{tabular}{p{3cm}p{2cm}R{5cm}R{5cm}}
\hline
\hline
& \textbf{Para.} & \textbf{Latent Class $1$} & \textbf{Latent Class $2$} \\
\hline
\hline
\multirow{4}{*}{\textbf{Mean}} 
& $\mu_{\eta_{0}}$ & $0.432$ ($0.243$, $0.892$) & $0.350$ ($0.200$, $0.707$) \\
& $\mu_{\eta_{1}}$ & $0.106$ ($0.053$, $0.294$) & $0.074$ ($0.042$, $0.174$) \\
& $\mu_{\eta_{2}}$ & $0.103$ ($0.052$, $0.280$) & $0.079$ ($0.042$, $0.164$) \\
& $\mu_{\gamma}$ & $0.055$ ($0.024$, $0.167$) & $0.062$ ($0.024$, $0.198$) \\
\hline
\hline
\multirow{4}{2.2cm}{\textbf{Variance}} 
& $\psi_{00}$ & $2.652$ ($1.731$, $4.817$) & $2.201$ ($1.400$, $3.789$) \\
& $\psi_{11}$ & $0.123$ ($0.069$, $0.272$) & $0.092$ ($0.057$, $0.170$) \\
& $\psi_{22}$ & $0.128$ ($0.071$, $0.333$) & $0.101$ ($0.062$, $0.219$) \\
\hline
\hline
\multirow{4}{2.2cm}{\textbf{Path Coef.}} 
& $\beta_{0}$ & ---\tnote{1} & $0.182$ ($0.084$, $0.592$) \\
& $\beta_{1}$ & --- & $0.120$ ($0.079$, $0.186$) \\
& $\beta_{2}$ & --- & $0.124$ ($0.083$, $0.199$) \\
\hline
\hline
\end{tabular}
\label{tbl:c2_RMSE3}
\begin{tablenotes}
\small
\item[1] ---: when fitting the proposed model, we set the first latent class as the reference group; accordingly, the coefficients of that class do not exist.
\end{tablenotes}
\end{threeparttable}
\end{table}

\renewcommand\thefigure{\arabic{figure}}
\setcounter{figure}{0}

\FloatBarrier
\begin{figure}
\centering
\includegraphics[width=0.8\textwidth]{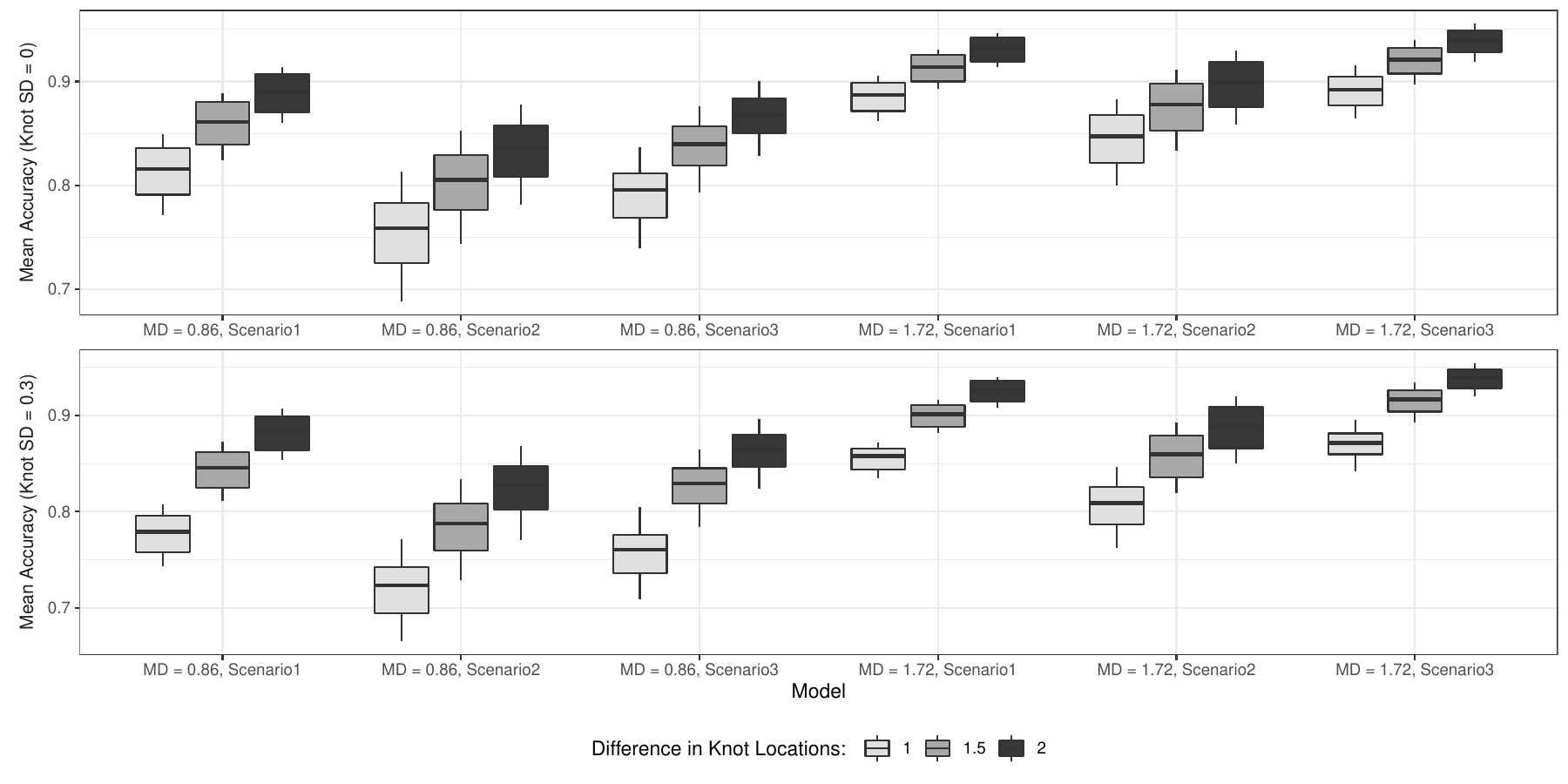}
\caption{Mean Accuracy of the Proposed Model across All Conditions with Two Latent Classes}
\label{fig:Accuracy}
\end{figure}

\begin{figure}
\centering
\includegraphics[width=0.8\textwidth]{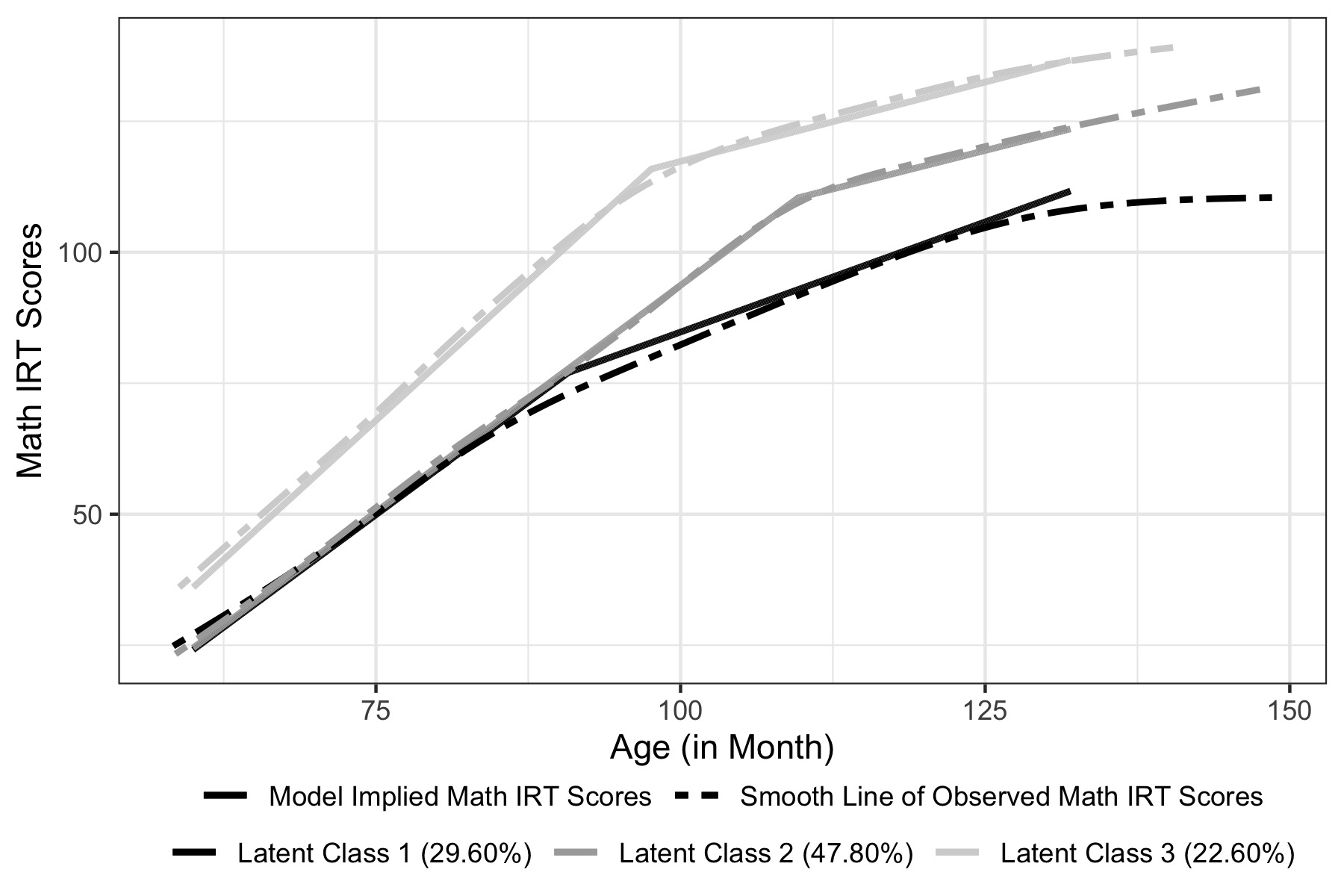}
\caption{Three Latent Classes: Model Implied Trajectories and Smooth Lines of Observed Mathematics IRT Scores}
\label{fig:math_curve}
\end{figure}

\begin{figure}
\centering
\includegraphics[width=0.8\textwidth]{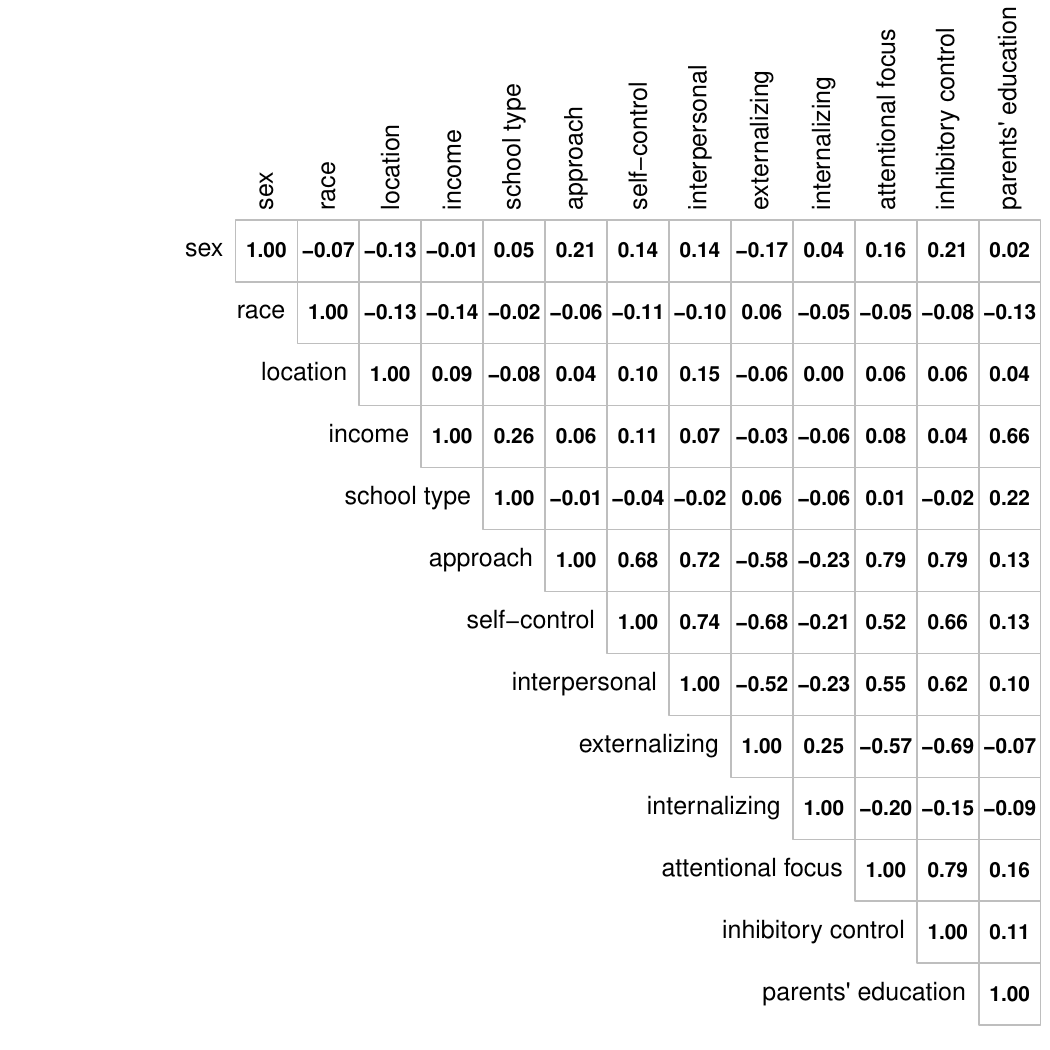}
\caption{Pairwise Correlation between Baseline Characteristics}
\label{fig:correlation}
\end{figure}

\renewcommand\thefigure{A.\arabic{figure}}
\setcounter{figure}{0}

\begin{figure}
\centering
\includegraphics[width=1.0\textwidth]{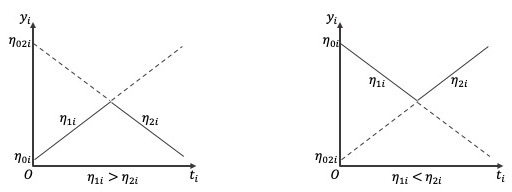}
\caption{Reparameterizing growth factors for Estimating a Fixed Knot}
\label{fig:proj1_2cases}
\end{figure}

\end{document}